\documentclass[aps,preprint,nofootinbib,preprintnumbers]{revtex4}
\usepackage{epsf,amssymb,amsfonts,amsmath,feynmf,array}

\def\bfnabla{\mbox{\boldmath $\nabla$}}

\def\bfsigma{\mbox{\boldmath $\sigma$}}
\def\bflambda{\mbox{\boldmath $\lambda$}}
\def\bfrho{\mbox{\boldmath $\rho$}}

\def\lQ{\Lambda_{\rm QCD}}

\def\als{\alpha_{\rm s}}
\newcommand{\MS}{\overline{\rm MS}}
\def\siml{{\ \lower-1.2pt\vbox{\hbox{\rlap{$<$}\lower6pt\vbox{\hbox{$\sim$}}}}\ }} 
\def\simg{{\ \lower-1.2pt\vbox{\hbox{\rlap{$>$}\lower6pt\vbox{\hbox{$\sim$}}}}\ }}

\def\vbfD{{\ \lower-8pt\vbox{\hbox{\rlap{$\!\leftrightarrow$}\lower8pt\vbox{\hbox{$\!\bf D$}}}}\ }} 
\def\dsl{\,\raise.15ex\hbox{/}\mkern-13.5mu D}

\newcommand{\nn}{\nonumber}
\newcommand{\be}{\begin{equation}} 
\newcommand{\ee}{\end{equation}}
\newcommand{\bea}{\begin{eqnarray}} 
\newcommand{\eea}{\end{eqnarray}}
\DeclareMathOperator{\Pathorder}{P}
 
\newcommand{\transpose}{\ensuremath{\text{T}}}
\newcommand{\bra}[1]{\ensuremath{\langle#1|}}
\newcommand{\ket}[1]{\ensuremath{|#1\rangle}}
\newcommand{\erw}[1]{\ensuremath{\langle#1\rangle}}
\newcommand{\erwzero}[1]{\ensuremath{\bra{0}#1\ket{0}}}

\renewcommand{\tensor}[1]{\ensuremath{\underline{\mathbf{#1}}}}
\newcommand{\Tensor}[1]{\ensuremath{\underline{\boldsymbol{#1}}}}
\newlength{\fillupPatternLength} 
\newlength{\fillupTextLength}

\newcommand{\Appendix}[1]%
    {%
     \section{#1}%
      }

\begin{document}
\begin{fmffile}{baryons}
\tolerance=10000 \hfuzz=5 pt
\baselineskip=24 pt 
\preprint{IFUM-808-FT} 
\title{Effective Field Theory Lagrangians for Baryons with Two and Three Heavy Quarks}

\vskip 0.25cm 

\author{Nora Brambilla}\email{nora.brambilla@mi.infn.it} 
\affiliation{Dipartimento di Fisica, INFN and Universit\`a degli Studi di Milano 
         via Celoria 16, 20133 Milano, Italy} 
\author{Thomas R\"osch}\email{t.roesch@thphys.uni-heidelberg.de, thomas.roesch@sap.com} 
\affiliation{Institut f\"ur Theoretische Physik, Universit\"at Heidelberg, 
         Philosophenweg 16, 69120 Heidelberg, Germany} 
\author{Antonio  Vairo}\email{antonio.vairo@mi.infn.it} 
\affiliation{Dipartimento di Fisica, INFN and Universit\`a degli Studi di Milano 
         via Celoria 16, 20133 Milano, Italy} 
\begin{abstract}
\noindent
By analogy with pNRQCD, we construct Effective Field Theories suitable 
to describe the heavy-quark sector of baryons made of two and three heavy quarks. 
A long-standing discrepancy between the hyperfine splitting 
of doubly heavy baryons obtained in the HQET and potential models is solved.
The one-loop matching of the 4-quark operators of dimension 6 is provided.
\end{abstract}
\maketitle

\vfill\eject
\section{Introduction}
The SELEX collaboration at Fermilab recently reported evidence of five resonances that 
may possibly be identified with doubly charmed baryon states  \cite{Ocherashvili:2004hi}. 
Tentatively the states have been interpreted as  
$ccd^+(3443)$, $ccd^+(3520)$, $ccu^{++}(3460)$, $ccu^{++}(3541)$ and $ccu^{++}(3780)$. 
Subsequently the $ccd^+(3520)$ state has been confirmed in two different decay modes
($\Xi_{cc}^{+} \rightarrow \Lambda_c^{+} K^{-} \pi^{+}$; $\Xi_{cc}^{+} \rightarrow  p D^{+} K^{-}$)
at a mass of  $3518.7 \pm 1.7$ MeV with an average lifetime less than 33 fs.
Although these findings need to be confirmed by other experiments and larger statistical samples,
they have triggered a renewed theoretical interest in doubly heavy baryon systems.

Doubly heavy baryons have been studied with several methods, mostly 
non-relativistic potential models (for some reviews see \cite{Fleck:mb,Kiselev:2001fw}), but  
also relativistic models \cite{Ebert:2002ig}, sum rules \cite{Bagan:1994dy,Kiselev:2000jb}
and in a chiral Lagrangian framework \cite{Bardeen:2003kt}. 
Masses of the lowest lying resonances have been obtained from lattice
calculations \cite{AliKhan:1999yb,Lewis:2001iz,Lewis:2001fy,Mathur:2002ce,Flynn:2003vz}.
Doubly heavy baryons are also suited to be studied in a QCD Effective Field Theory (EFT) framework. 
Indeed, they are characterized by at least two widely separated scales: the large heavy-quark masses, $m$,  
and the low momentum transfer between the heavy and the light quarks, which is of order $\lQ$.
If one assumes that the typical momentum transfer between the 
two heavy quarks is larger than $\lQ$, then a $QQq$ baryon is very similar 
to a bound state of a heavy antiquark and a light quark.
This has first been noted in \cite{Savage:di}, where at leading order in
$\lQ/m$  the hyperfine splitting of the doubly heavy baryon ground state 
has been related to the ground-state hyperfine splitting 
of the heavy-light meson. In \cite{Savage:pr} non-leptonic and semileptonic decays 
of doubly heavy baryons have been examined in the context of $SU(3)$ flavour symmetry.  
After this original work no further step has been made in the 
direction of providing a systematic description of doubly heavy baryons 
in an EFT framework that fully combines the dynamics of the 
two heavy quarks with that one of the light one. Following some suggestions in \cite{Soto:2003ft}, 
with this work we attempt to make such further step. In particular, we identify the degrees of freedom 
and write the low-energy EFT Lagrangian that describes doubly heavy baryon systems in the heavy-quark 
sector, once the heavy-quark momentum-transfer scale has been integrated out. 
The framework is similar to that one developed in the last years for heavy-quarkonium 
systems (for a review see \cite{Brambilla:2004jw}). 

Baryons made of three heavy quarks $QQQ$  have not been observed yet. Their relevance 
has been emphasized since long ago \cite{bjorken}. They would reveal a pure baryonic spectrum 
without light-quark complications and provide valuable insight into the
quark confinement mechanism. Indeed, the three-quark static Wilson loop
is intensively studied on the lattice \cite{Bali:2000gf,Takahashi:2004rw}
as a source of information about the baryon heavy-quark potential and 
the type of confining configurations \cite{Brambilla:1995px,Kuzmenko:2002zs,Cornwall:2003vn}. 
In this work we will identify the degrees of freedom and write the low-energy EFT Lagrangian
that describes heavy baryons made of three heavy quarks, once the heavy-quark 
momentum-transfer scale has been integrated out. We will express the leading-order 
and spin-dependent potentials in terms of Wilson loop amplitudes along the lines developed 
for heavy quarkonia in \cite{Brambilla:2000gk}.

A recent review that also discusses the present status of the art, experimental and theoretical, including 
lattice, for heavy baryons made with two or three heavy quarks is Ref.~\cite{Brambilla:2004wf}. 
We refer to it for a more complete bibliography on the subject. 
This work is partially based on \cite{Thomas}. We refer to it for details in
some of the derivations.

The paper is distributed as follows. In Sec.~\ref{secNRQCD} we introduce NRQCD for heavy baryons. 
In Sec.~\ref{secQQq} we write the low-energy EFT for $QQq$ baryons and 
calculate the hyperfine splitting of the ground state. In Sec.~\ref{secQQQ} 
we write the low-energy EFT for $QQQ$ baryons and give some exact
non-perturbative expressions for the leading-order and spin-dependent potentials. 
Sec.~\ref{secCON} is devoted to the conclusions. 
Some technical details may be found in the appendices.

\section{NRQCD}
\label{secNRQCD}
Non-relativistic QCD (NRQCD) is the EFT suitable to describe 
systems made of two or more heavy quarks. It is obtained from QCD 
by integrating out modes of energy of the order of the heavy-quark masses 
\cite{Caswell:1985ui}. 

We are interested here only in the heavy-quark sector of the NRQCD Lagrangian. 
The 2-heavy-quark sector coincides with the Lagrangian of the Heavy Quark 
Effective Theory (HQET). Up to order $1/m^2$ it reads:
\bea
{\cal L}^{\rm NRQCD}_{Q}= 
\sum_{h=1}^{N_Q}
Q_h^\dagger \left[ i D_0 + \, \frac{{\boldsymbol{D}}^2}{2 m_h}
+ c^{(h)}_F\,\frac{\bfsigma \cdot g{\boldsymbol{B}}}{2 m_h}
+ c^{(h)}_D \, \frac{ \left[{\boldsymbol{D}} \cdot, g {\boldsymbol{E}} \right]}{8 m^2_h}
+ i c^{(h)}_S \, \frac{ \bfsigma \cdot \left[{\boldsymbol{D}} \times, g {\boldsymbol{E}} \right]}{8 m^2_h} \right] Q_h, 
\label{lNRQCDQ}
\eea
where $N_Q$ is the number of heavy-quark flavours, 
$Q_h$ the Pauli spinor field that annihilates the quark of flavour $h$ and mass $m_h$, 
$i D_0=i\partial_0 -gA^0$, $i{\boldsymbol{D}}=i\bfnabla+g{\boldsymbol{A}}$,
$[{\boldsymbol{D} \cdot, \boldsymbol{E}}]={\boldsymbol{D} \cdot \boldsymbol{E}} - {\boldsymbol{E} \cdot \boldsymbol{D}}$, 
$[{\boldsymbol{D} \times, \boldsymbol{E}}]={\boldsymbol{D} \times \boldsymbol{E} -\boldsymbol{E} \times \boldsymbol{D}}$, 
${\boldsymbol{E}}^i = F^{i0}$, ${\boldsymbol{B}}^i = -\epsilon_{ijk}F^{jk}/2$ ($\epsilon_{123} = 1$) and 
$\bfsigma = (\bfsigma_1, \bfsigma_2, \bfsigma_3)$ are the Pauli matrices.
The matching coefficients $c^{(h)}_F$, $c^{(h)}_D$ and $c^{(h)}_S$ 
may be found at one loop, for instance, in \cite{Manohar:1997qy}.
The Lagrangian (\ref{lNRQCDQ}) (with ${\cal O}(1/m^3)$ terms included, but   
all matching coefficients set equal to 1) has been used to perform lattice 
calculations of the spectra of heavy baryons in \cite{Mathur:2002ce}. 

At order $1/m^2$ the NRQCD Lagrangian relevant to describe 
baryons made of two or more heavy quarks exhibits also a 4-heavy-quark sector: 
\bea
{\cal L}^{\rm NRQCD}_{QQ} &=&
\sum_{h'\ge h =1}^{N_Q} \left(
  \frac{d^{ss}_{Q_hQ_{h'}}}{m_h m_{h'}} Q_h^{\dag} Q_h Q_{h'}^{\dag} Q_{h'}
+ \frac{d^{sv}_{Q_hQ_{h'}}}{m_h m_{h'}} Q_h^{\dag} {\bfsigma} Q_h \cdot Q_{h'}^{\dag} {\bfsigma} Q_{h'}
\right.
\nn
\\
&&
\left.
+ \frac{d^{vs}_{Q_hQ_{h'}}}{m_h m_{h'}} 
\sum_{a=1}^8 Q_h^{\dag} {\rm T}^a Q_h Q_{h'}^{\dag} {\rm T}^a Q_{h'}
+ \frac{d^{vv}_{Q_hQ_{h'}}}{m_h m_{h'}} 
\sum_{a=1}^8 Q_h^{\dag} {\rm T}^a {\bfsigma} Q_h \cdot Q_{h'}^{\dag} {\rm T}^a {\bfsigma} Q_{h'}
\right)
\,.
\label{lNRQCDQQ}
\eea
The matching coefficients $d_{Q_hQ_{h'}}$ start getting contributions at order
$\als^2$ . They have been calculated to this order in appendix \ref{appA}.

Six-quark operators contribute to heavy baryons made of three heavy quarks. They show up at order $1/m^5$.
The corresponding matching coefficients start getting contributions at order $\als^4$. 
Hence, these operators are highly suppressed and will be neglected in the rest of the paper.

In the following we will make a step further and construct the EFT suitable 
to describe baryons made of two (Sec.~\ref{secQQq}) and three
(Sec.~\ref{secQQQ}) heavy quarks once gluons of energy or momentum of the 
order of the momentum transfer between the heavy quarks have been integrated
out. The procedure and the resulting EFT will be quite similar to that one developed 
for heavy quarkonium in \cite{Pineda:1997bj,Brambilla:1999xf}. For this reason we will
call the EFT with the same name: potential NRQCD (pNRQCD).

\section{pNRQCD for $QQq$  baryons}
\label{secQQq}
In this section we deal with baryons made of two heavy quarks $Q_1$, $Q_2$
($bb$, $bc$ or $cc$) with masses $m_1$ and $m_2$ respectively and one massless 
quark $q$. The dynamics of these systems is expected to mix 
aspects typical of heavy quarkonium with aspects typical of heavy-light
mesons. On the one hand, the interaction of the two heavy quarks is that one 
of a non-relativistic quark pair close to threshold moving with relative 
velocity $v$. It is, therefore, characterized by the energy scales: $m \gg mv
\gg mv^2$, where $mv$ is the scale of the typical momentum transfer 
between the two heavy quarks (or of the inverse of their typical distance) 
and $mv^2$ is the typical binding energy.
On the other hand, the energy scale that governs the interaction
between the heavy quarks and the light one is $\lQ$.
Two different situations are possible. 

{\bf (A)} If $mv \gg \lQ$, at a scale $\mu$ such that $mv \gg \mu \gg \lQ$ 
the heavy-quark distance cannot be resolved.
The $Q_1Q_2$ pair behaves like a point-like particle (sometimes also called 
diquark \cite{Anselmino:1992vg}) in an antitriplet or sextet colour configuration. 
In the antitriplet configuration the two heavy quarks attract each other. 
The interaction of the antitriplet field with the light quark is similar to that one of 
a heavy antiquark with a light quark in a $D$ or $B$ meson \cite{Savage:di}.
However, the spectrum  is expected to be richer and more complex 
due to the internal excitations of the heavy-quark system. 
These include radial and spin excitations, but also colour excitations to sextet  
configurations. Considering that the scale $\mu$ is perturbative, 
in this situation pNRQCD has the following degrees of freedom:
light quarks, gluons of energy and momentum lower than $mv$ (also called
ultrasoft) and heavy quarks. To ensure that the gluons are of energy and
momentum lower than $mv$, gluons appearing in vertices involving heavy-quark
fields are multipole expanded in the relative distance $r \sim 1/(mv)$ between the two heavy
quarks. This corresponds to expanding in $r\lQ \sim \lQ/(mv)$ or $r\,(mv^2) \sim v$.
In the situation $mv \gg \lQ$, in principle one may further distinguish between the subcases $mv^2 \gg \lQ$, 
$mv^2 \sim \lQ$  and $\lQ \gg mv^2$. In the first case, one may expand in $\lQ/(mv^2)$ and 
disentangle the heavy-heavy dynamics, which is completely accessible to perturbation theory,
from the heavy-light one. In general, excitations of the heavy-heavy system will
dominate over excitations of the heavy-light system. In the latter case, the 
potential governing the heavy-heavy system gets non-perturbative contributions.\footnote{
An analogous situation for the quarkonium system has been treated in \cite{Brambilla:1999xf}.}
Since the kinetic energy of the heavy quarks is smaller than $\lQ$, the heavy-light 
dynamics dominates in this situation over the heavy-heavy one. At leading 
order in the $mv^2/\lQ$ expansion the flavour symmetry typical of the HQET is restored.

In the rest of this section, we will deal with the general situation  $mv \gg
\lQ$, without assuming any special hierarchy between the scales $mv^2$ and $\lQ$. 
To be definite, one may think that we work in the situation $mv^2 \sim \lQ$. 

{\bf (B)} If $mv \sim \lQ$ the distances between the three quarks are of the same magnitude.
Hence, we cannot disentangle the heavy-quark pair dynamics from the light-quark one.
Moreover, the potential between the two heavy quarks is non-perturbative. 
At the level of NRQCD, the system may be studied with lattice calculations. 
In this situation it seems unlikely that a simple diquark--light-quark picture holds.
In general, from an EFT point of view it does not seem consistent to have 
a diquark--light-quark picture for the heavy-quarks--light-quark interaction, 
which implicitly assumes $mv \gg \lQ$, and at the same time a confining  
potential binding the two heavy quarks, which requires $mv \sim \lQ$, 
as so often done in potential models.

In the following, we will work out pNRQCD in the 
situation {\bf (A)}. This situation is expected to be appropriate 
for the description of at least doubly heavy baryons in the ground state.

\subsection{Lagrangian}
In this section we write the pNRQCD Lagrangian that describes  
heavy baryons of the type $Q_1Q_2q$ in the situation where 
the typical momentum transfer between the two heavy 
quarks is much larger than $\lQ$. This corresponds to 
the case labeled {\bf (A)} above.

The number of allowed operators is reduced if we
choose to have a manifestly gauge invariant Lagrangian. This may be obtained by 
projecting the Lagrangian on the heavy-heavy sector of the Fock space,  
by splitting the heavy-heavy fields into an antitriplet and sextet component
(${\boldsymbol{r}} = {\boldsymbol{x}}_1 - {\boldsymbol{x}}_2$, 
${\boldsymbol{R}} = (m_1{\boldsymbol{x}}_1 + m_2{\boldsymbol{x}}_2)/(m_1+m_2)$), 
\be 
Q_{1\,i}({\boldsymbol{x}}_1,t) Q_{2\,j}({\boldsymbol{x}}_2,t) 
\sim \sum_{\ell =1}^3 T^\ell({\boldsymbol{r}},{\boldsymbol{R}},t) \, \tensor{T}^\ell_{ij}
+ \sum_{\sigma =1}^6 \Sigma^\sigma({\boldsymbol{r}},{\boldsymbol{R}},t) \, \Tensor{\Sigma}^\sigma_{ij}, \quad
i,j=1,2,3\,, 
\label{QQTS}
\ee
and by building the Lagrangian from these operators.
The tensors $\tensor{T}^\ell_{ij}$ and $\Tensor{\Sigma}^\sigma_{ij}$ are 
defined in appendix \ref{App:Multiplet_TensorsQQ}.
The Lagrangian is constrained to satisfy all the symmetries 
of QCD. In particular, it is symmetric under the exchange of the heavy quarks.
Such symmetry transformation changes $m_1 \leftrightarrow m_2$ 
and ${\boldsymbol{r}}$ to $-{\boldsymbol{r}}$. The gluon fields are even, because multipole 
expanded around the centre-of-mass of the heavy-heavy system.
For what concerns the heavy-quark fields, from Eq.~(\ref{QQTS}) it follows that 
$T^\ell$ is even, because $\tensor{T}^\ell_{ij}$ is odd under the exchange 
$i \leftrightarrow j$, and $\Sigma^\sigma$ is odd, because 
$\Tensor{\Sigma}^\sigma_{ij}$ is even under the exchange $i \leftrightarrow j$. 

The resulting Lagrangian $\mathcal{L}_\text{pNRQCD} = \mathcal{L}_\text{pNRQCD}({\boldsymbol{R}},t)$
at ${\cal O}(1/m)$ in the $1/m$ expansion and at ${\cal O}(r)$ in the multipole expansion is 
\begin{equation}
\mathcal{L}_\text{pNRQCD} = 
\mathcal{L}_\text{gluon} 
+ \mathcal{L}_\text{light} 
+ \mathcal{L}^{(0,0)}_\text{pNRQCD} 
+ \mathcal{L}^{(0,1)}_\text{pNRQCD} 
+ \mathcal{L}^{(1,0)}_\text{pNRQCD}, 
\label{qqpl}
\end{equation}
with
\bea
&&\mathcal{L}_\text{gluon} = -\frac{1}{4} \sum_{a=1}^8 F^a_{\mu\nu}F^{a\mu\nu},
\label{Lgluon}\\
&& \nn\\
&&\mathcal{L}_\text{light}  =  \sum_{f=1}^3 \bar{q}_f \, i \dsl \,  q_f,
\label{Llight}
\\
&& \nn\\
\nn
\eea
\bea
&&\mathcal{L}^{(0,0)}_\text{pNRQCD} = 
\int d^3r \, T^\dagger \left[ iD_0 - V_T^{(0)} \right] T 
+ \Sigma^\dagger \left[ iD_0 - V^{(0)}_\Sigma \right] \Sigma,
\label{qqpl0}
\\
&& \nn\\
&&\mathcal{L}^{(0,1)}_\text{pNRQCD} =
-\int d^3r \, V_{T{\boldsymbol{r}}\cdot{\boldsymbol{E}}\Sigma}^{(0,1)}
\sum_{a=1}^8\sum_{\ell=1}^3\sum_{\sigma=1}^6
\left[ 
\left( \sum_{ijk=1}^3 \tensor{T}^\ell_{ij}T^a_{jk} \Tensor{\Sigma}^{\sigma}_{ki} \right)\;
T^{\ell\dagger} \boldsymbol{r} \cdot g\boldsymbol{E}^a \,\Sigma^\sigma 
\right.
\nn\\
&&\hspace{30mm}
\left.
- \left( \sum_{ijk=1}^3 \Tensor{\Sigma}^\sigma_{ij} T^a_{jk} \tensor{T}^{\ell}_{ki} \right)\;
\Sigma^{\sigma\dagger} \boldsymbol{r} \cdot g\boldsymbol{E}^a \,T^\ell \right] 
\nn\\
&& \hspace{3mm}
- \frac{m_1-m_2}{2m_R} V_{T{\boldsymbol{r}}\cdot{\boldsymbol{E}}T}^{(0,1)} \, 
\sum_{a=1}^8
T^\dagger \boldsymbol{r} \cdot g \boldsymbol{E}^a T^a_{\bar{3}} \,T 
- \frac{m_1-m_2}{2m_R} V_{\Sigma{\boldsymbol{r}}\cdot{\boldsymbol{E}}\Sigma}^{(0,1)} \, 
\sum_{a=1}^8
\Sigma^\dagger \boldsymbol{r} \cdot g \boldsymbol{E}^a T^a_{6} \,\Sigma, 
\label{qqpl01}
\\
&& \nn\\
&&\mathcal{L}^{(1,0)}_\text{pNRQCD} =
\int d^3r \, T^\dagger \left[ \frac{\boldsymbol{D}_R^2}{2m_R} 
+ \frac{\boldsymbol{\nabla}_r^2}{2m_r} 
\right] T 
+ \Sigma^\dagger \left[ \frac{\boldsymbol{D}_R^2}{2m_R} 
+ \frac{\boldsymbol{\nabla}_r^2}{2m_r} 
\right] \Sigma
\nn\\
&& \hspace{3mm}
+ V_{T\bfsigma\cdot{\boldsymbol{B}}\Sigma}^{(1,0)}
\sum_{a=1}^8\sum_{\ell=1}^3\sum_{\sigma=1}^6
\left[ 
\left( \sum_{ijk=1}^3 \tensor{T}^\ell_{ij}T^a_{jk} \Tensor{\Sigma}^{\sigma}_{ki} \right)\;
 T^{\ell\dagger} \left( - \frac{c^{(1)}_F\bfsigma^{(1)}}{2m_1} +
\frac{c^{(2)}_F\bfsigma^{(2)}}{2m_2} \right) \cdot g{\boldsymbol{B}}^a \,\Sigma^\sigma 
\right.
\nn\\
&&\hspace{30mm}
\left.
- \left( \sum_{ijk=1}^3 \Tensor{\Sigma}^\sigma_{ij} T^a_{jk} \tensor{T}^{\ell}_{ki} \right)\;
\Sigma^{\sigma\dagger} \left( - \frac{c^{(1)}_F\bfsigma^{(1)}}{2m_1} +
\frac{c^{(2)}_F\bfsigma^{(2)}}{2m_2} \right) \cdot g{\boldsymbol{B}}^a \, T^\ell 
\right] 
\nn\\
&&\hspace{3mm}
+ \frac{V_{T\bfsigma\cdot{\boldsymbol{B}}T}^{(1,0)}}{2}
\sum_{a=1}^8
T^\dagger \left(\frac{c^{(1)}_F\bfsigma^{(1)}}{2m_1} 
+ \frac{c^{(2)}_F\bfsigma^{(2)}}{2m_2} \right) \cdot g{\boldsymbol{B}}^a  T^a_{\bar{3}}\, T 
\nn\\
&&\hspace{3mm}
+ \frac{V_{\Sigma\bfsigma\cdot{\boldsymbol{B}}\Sigma}^{(1,0)}}{2} 
\sum_{a=1}^8
\Sigma^\dagger \left(\frac{c^{(1)}_F\bfsigma^{(1)}}{2m_1} +
\frac{c^{(2)}_F\bfsigma^{(2)}}{2m_2} \right) \cdot g{\boldsymbol{B}}^a T^a_{6} \, \Sigma
\nn\\
&&\hspace{3mm}
+ V_{T{\bf L}\cdot{\boldsymbol{B}}\Sigma}^{(1,0)} \frac{m_1-m_2}{2m_Rm_r} 
\sum_{a=1}^8\sum_{\ell=1}^3\sum_{\sigma=1}^6
\left[ 
\left( \sum_{ijk=1}^3 \tensor{T}^\ell_{ij}T^a_{jk} \Tensor{\Sigma}^{\sigma}_{ki} \right)\;
T^{\ell\dagger} {\bf L}_r\cdot g{\boldsymbol{B}}^a \, \Sigma^\sigma
\right.
\nn\\
&&\hspace{30mm}
\left.
-\left(\sum_{ijk=1}^3 \Tensor{\Sigma}^\sigma_{ij} T^a_{jk} \tensor{T}^{\ell}_{ki} \right)\;
\Sigma^{\sigma\dagger} {\bf L}_r\cdot g{\boldsymbol{B}}^a \, T^\ell 
\right] 
\nn\\
&&\hspace{3mm}
+ \frac{V_{T{\bf L}\cdot{\boldsymbol{B}}T}^{(1,0)}}{4} 
\left(\frac{1}{m_r}- \frac{2}{m_R}\right)
\sum_{a=1}^8
T^\dagger {\bf L}_r \cdot g{\boldsymbol{B}}^a  T^a_{\bar{3}} \, T 
\nn\\
&&\hspace{3mm}
+\frac{V_{\Sigma{\bf L}\cdot{\boldsymbol{B}}\Sigma}^{(1,0)}}{4} 
\left(\frac{1}{m_r} - \frac{2}{m_R}\right)
\sum_{a=1}^8
\Sigma^\dagger {\bf L}_r \cdot g{\boldsymbol{B}}^a  T^a_{6}\, \Sigma ,
\label{qqpl10}
\eea
where $m_R = m_1+m_2$, $m_r = m_1m_2/(m_1+m_2)$, $\bfsigma^{(h)}$ is the Pauli 
matrix acting on the heavy quark $h$, $i{\boldsymbol{D}}_R = i\bfnabla_R+g{\boldsymbol{A}}$,
${\bf L}_r =  {\boldsymbol{r}}\times (-i\bfnabla_r)$, $T = (T^1,T^2,T^3)$, 
$\Sigma = (\Sigma^1, \Sigma^2, ..., \Sigma^6)$, the gauge fields in the
covariant derivatives acting on the antitriplet and sextet are understood in the 
antitriplet and sextet representation respectively, $T^a_{\bar{3}}$ and 
$T^a_{6}$ have been defined in appendix \ref{App:Group_rep} and all gluon fields are
evaluated in $({\boldsymbol{R}},t)$. 
The coefficients $c^{(1)}_F$ and $c^{(2)}_F$ are the Wilson coefficients of
NRQCD introduced in Sec.~\ref{secNRQCD}. The functions $V$ are the Wilson
coefficients of pNRQCD for doubly heavy baryons. They encode the contributions 
coming from gluons of energy or momentum of order $mv$, which have been integrated out. 
They are non-analytic functions of $r$. As we will discuss in the next section,  
at tree level they are
\bea 
V_{T{\boldsymbol{r}}\cdot{\boldsymbol{E}}\Sigma}^{(0,1)} = V_{T{\boldsymbol{r}}\cdot{\boldsymbol{E}}T}^{(0,1)}= 
V_{\Sigma{\boldsymbol{r}}\cdot{\boldsymbol{E}}\Sigma}^{(0,1)}= 1,
\nn\\
V_{T\bfsigma\cdot{\boldsymbol{B}}\Sigma}^{(1,0)} = V_{T\bfsigma\cdot{\boldsymbol{B}}T}^{(1,0)} = 
V_{\Sigma\bfsigma\cdot{\boldsymbol{B}}\Sigma}^{(1,0)}=1,
\label{QQtree}\\
V_{T{\bf L}\cdot{\boldsymbol{B}}\Sigma}^{(1,0)} = V_{T{\bf L}\cdot{\boldsymbol{B}}T}^{(1,0)}= 
V_{\Sigma{\bf L}\cdot{\boldsymbol{B}}\Sigma}^{(1,0)}= 1,
\nn
\eea
while $V^{(0)}_T$ and $V^{(0)}_\Sigma$ get the first non-vanishing contribution at order $\als$. 
In Eqs.~(\ref{qqpl01}) and (\ref{qqpl10}) we have displayed only the operators 
that have a non-vanishing tree-level matching coefficient. 
The coefficients in front of the $\boldsymbol{D}_R^2$ and $\bfnabla^2_r$
operators in (\ref{qqpl10}) are equal to 1, due to Poincar\'e invariance or
dynamical considerations similar to those developed in
\cite{Brambilla:2003nt}.
We observe that in the case $m_1\neq m_2$, 
electric dipole transitions between antitriplet states induced by the term
\be
-\frac{m_1-m_2}{2m_R} \sum_{a=1}^8 
T^\dagger \boldsymbol{r} \cdot g \boldsymbol{E}^a  T^a_{\bar{3}} \,T 
\ee
are allowed \cite{Kiselev:2001fw,Gershtein:nx}.

The power counting of the Lagrangian (\ref{qqpl}) in the centre-of-mass frame goes as follows: 
$\bfnabla_r \sim mv$, ${\boldsymbol{r}} \sim 1/(mv)$, ${\boldsymbol{D}}_{R} \sim \lQ$, $mv^2$, 
$V^{(0)}_{T,\Sigma} \sim mv^2$ and ${\boldsymbol{E}},{\boldsymbol{B}} \sim \lQ^2$, $(mv^2)^2$.
The power counting is not unique, because the scales $mv^2$ and $\lQ$ are
still entangled in the dynamics. The Lagrangian at leading order reads
\bea
\mathcal{L}^\text{LO}_\text{pNRQCD} &=& 
\int d^3r \, T^\dagger \left[ iD_0 + \frac{\boldsymbol{\nabla}_r^2}{2m_r} - V_T^{(0)} \right] T 
+ \Sigma^\dagger \left[ iD_0 + \frac{\boldsymbol{\nabla}_r^2}{2m_r} - V^{(0)}_\Sigma \right] \Sigma
\nn\\
&& -\frac{1}{4}  \sum_{a=1}^8 F^a_{\mu\nu}F^{a\mu\nu}
+ \sum_{f=1}^3 \bar{q}_f \, i \dsl \,  q_f.
\label{qqLO}
\eea

\subsection{Matching}
The matching from NRQCD to pNRQCD is, in general, performed by 
calculating Green functions in the two theories and imposing that 
they are equal order by order in the inverse of the mass and in the multipole expansion.
Since we are working in the  situation where the typical momentum transfer between 
the heavy quarks is larger than $\lQ$, we can, in addition,
perform the matching order by order in $\als$.

If we aim at calculating the matching at tree level a convenient  
approach consists in projecting the NRQCD Hamiltonian on the two-quark Fock space spanned by 
\be
\int d^3x_1\,d^3x_2 \, \sum_{ij=1}^{3}
\Phi_{Q_1Q_2}^{ij}(\boldsymbol{x}_1,
\boldsymbol{x}_2) Q^{i \,\dagger}_1 (\boldsymbol{x}_1)
Q^{j\,\dagger}_2 (\boldsymbol{x}_2) \ket{{0}},
\label{FockQQ}
\ee
where $\ket{0}$ is the Fock subspace containing no heavy quarks 
but  an arbitrary number of ultrasoft gluons and light quarks and 
$\Phi_{Q_1Q_2}(\boldsymbol{x}_1,\boldsymbol{x}_2)$ is a $3 \otimes 3$ tensor
in colour space and a $2 \otimes 2$ tensor in spin space. 
This is similar to what is done in \cite{Pineda:1998kn}.
After projection, all gluon fields are multipole expanded in ${\boldsymbol{r}}$.
In order to make gauge invariance explicit at the Lagrangian level,   
it is useful to decompose $\Phi_{Q_1Q_2}(\boldsymbol{x}_1,\boldsymbol{x}_2)$ into a 
field $T(\boldsymbol{r}, \boldsymbol{R}, t)$, which transforms like a colour antitriplet, 
and a field $\Sigma(\boldsymbol{r}, \boldsymbol{R}, t)$, which transforms like
a colour sextet:
\bea
\Phi_{Q_1Q_2}^{ij} (\boldsymbol{x}_1, \boldsymbol{x}_2, t) &=& \sum_{i'j'=1}^{3}
\phi_{ii'} (\boldsymbol{x}_1, \boldsymbol{R}, t) \,
\phi_{jj'} (\boldsymbol{x}_2, \boldsymbol{R}, t) \,
\nn\\
&&\quad \times 
\left( \sum_{\ell=1}^3 T^\ell(\boldsymbol{r}, \boldsymbol{R}, t) \, \tensor{T}^\ell_{i'j'} +
\sum_{\sigma=1}^6 \Sigma^\sigma(\boldsymbol{r}, \boldsymbol{R}, t) \, \Tensor{\Sigma}^\sigma_{i'j'} \right),
\label{stateQQ}
\eea
where 
\be
\phi(\boldsymbol{y}, \boldsymbol{x}, t) \equiv \Pathorder \exp \left(
ig \int_0^1 ds \, ( \boldsymbol{y} - \boldsymbol{x} ) \cdot
\boldsymbol{A}(\boldsymbol{x} + ( \boldsymbol{y} -
\boldsymbol{x} )s,t) \right).
\label{WilsonString}
\ee
$\Pathorder$ stands for path ordering. At leading order in the coupling constant, 
$\phi_{ij} (\boldsymbol{x}_1, \boldsymbol{R}, t) = \delta_{ij}$ and 
\be
\Phi_{Q_1Q_2}^{ij} (\boldsymbol{x}_1, \boldsymbol{x}_2, t) \approx
\sum_{\ell=1}^3 T^\ell(\boldsymbol{r}, \boldsymbol{R}, t) \, \tensor{T}^\ell_{ij} +
\sum_{\sigma=1}^6 \Sigma^\sigma(\boldsymbol{r}, \boldsymbol{R}, t) \, \Tensor{\Sigma}^\sigma_{ij}.
\label{stateQQLO}
\ee
After projecting on (\ref{FockQQ}), one obtains the Lagrangian (\ref{qqpl0})-(\ref{qqpl10})
with the matching conditions (\ref{QQtree}).

As an example, let us consider the calculation that leads to the term 
\be
\delta \mathcal{L}_\text{pNRQCD} = 
\int d^3r\,
\frac{1}{2} 
\sum_{a=1}^8
T^\dagger \left(\frac{c^{(1)}_F\bfsigma^{(1)}}{2m_1} +
\frac{c^{(2)}_F\bfsigma^{(2)}}{2m_2} \right) \cdot g{\boldsymbol{B}}^aT^a_{\bar{3}} 
\, T 
\label{pnrqcdQQSB}
\ee
in the pNRQCD Lagrangian (see Eq.~(\ref{qqpl10})). 
We start from the NRQCD term
\be 
\delta \mathcal{L}_\text{NRQCD} = 
Q_1^\dagger c^{(1)}_F\, \frac{\bfsigma \cdot g {\boldsymbol{B}}}{ 2 m_1}Q_1
+ Q_2^\dagger c^{(2)}_F\, \frac{\bfsigma \cdot g {\boldsymbol{B}}}{2 m_2}Q_2\,.
\label{nrqcdQQSB}
\ee
Projecting onto (\ref{stateQQLO}), we obtain in the antitriplet-antitriplet sector
\bea
\delta \mathcal{L}_\text{pNRQCD} &=& 
\int d^3r 
\sum_{a=1}^8\sum_{\ell\ell'\,ijk=1}^3 
\left( T^{\ell\dagger} \tensor{T}^\ell_{ij}\,
\frac{c^{(1)}_F\bfsigma^{(1)}}{2m_1}\cdot g{\boldsymbol{B}}^aT^a_{ ik} \, 
T^{\ell'}\tensor{T}^{\ell'}_{kj}
\right.
\nn\\
&& \qquad\qquad\qquad\qquad\qquad\qquad
\left. +
T^{\ell\dagger} \tensor{T}^\ell_{ij} \,
\frac{c^{(2)}_F\bfsigma^{(2)}}{2m_2}\cdot g{\boldsymbol{B}}^aT^a_{ jk} \,
T^{\ell'} \tensor{T}^{\ell'}_{ik}\right).
\eea
Using the definition (\ref{Tabg}), we have
\bea
\sum_{ijk=1}^3 \tensor{T}^\ell_{ij} T^a_{ ik} \tensor{T}^{\ell'}_{kj}
&=& - \frac{T^a_{\ell'\ell}}{2} = \frac{(T^a_{\bar{3}})_{\ell\ell'}}{2}, 
\\
\sum_{ijk=1}^3 \tensor{T}^\ell_{ij}T^a_{ jk}\tensor{T}^{\ell'}_{ik}
&=& - \frac{T^a_{\ell'\ell}}{2} = \frac{(T^a_{\bar{3}})_{\ell\ell'}}{2}, 
\eea
and eventually end up with Eq.~(\ref{pnrqcdQQSB}). 
This fixes $V_{T\bfsigma\cdot{\boldsymbol{B}}T}^{(1,0)} = 1$ at leading order.
Note that Eq.~(\ref{pnrqcdQQSB}) differs by a factor $1/2$ from Eqs.~(9) and
(10) in \cite{Savage:di}, which seem to miss the correct colour normalization of the antitriplet states.

One may ask what happens to $V_{T\bfsigma\cdot{\boldsymbol{B}}T}^{(1,0)}$ beyond tree level.
Order $\als$ corrections may only come from one-gluon corrections to the 
NRQCD vertex of Eq.~(\ref{nrqcdQQSB}), because all other spin-dependent operators 
in NRQCD contribute to higher orders in $1/m$. One-loop corrections to the 
external (transverse) gluon or to a quark line or involving a gluon attached 
to the external gluon and to the quark line coupled to it 
vanish in dimensional regularization, once we have expanded in the external energies.
Gluons attached to a quark line are longitudinal. It is convenient to use the Coulomb gauge. 
In Coulomb gauge, longitudinal gluons exchanged between different quark lines  
cancel in the matching with equal contributions from the pNRQCD side. 
Finally, longitudinal gluons attached to the external gluon line and a 
heavy quark line not coupled to it contribute to higher-order operators, 
since the three-gluon vertex is proportional to the external energies. 
We conclude that $V_{T\bfsigma\cdot{\boldsymbol{B}}T}^{(1,0)}$ does not get contributions at one
loop, so that $V_{T\bfsigma\cdot{\boldsymbol{B}}T}^{(1,0)} =  1 + {\cal O}(\als^2)$.
Similar considerations hold for $V_{\Sigma\bfsigma\cdot{\boldsymbol{B}}\Sigma}^{(1,0)}$
and $V_{T\bfsigma\cdot{\boldsymbol{B}}\Sigma}^{(1,0)}$.

The perturbative matching of the static potentials $V^{(0)}_T$ and $V^{(0)}_\Sigma$ goes as
follows (see \cite{Brambilla:1999xf} for the quarkonium case). 
In NRQCD we compute static Green functions, whose initial and final states
overlap with the antitriplet and sextet fields in pNRQCD.
Since we work order by order in $\als$, it is not necessary for the Green
functions to be gauge invariant. A possible choice is
\bea
&&\hspace{-5mm} I^{uv}_\mathcal{M} \equiv \!\!\!
\sum_{iji'j'=1}^{3}
\bra{0} \Tensor{\mathcal{M}}^u_{ij} \, 
Q_i(\boldsymbol{R},\boldsymbol{x}_1,T/2) 
Q_j(\boldsymbol{R}, \boldsymbol{x}_2, T/2) \,
\Tensor{\mathcal{M}}^{v}_{i'j'} \,
Q^\dagger_{i'} (\boldsymbol{R}, \boldsymbol{y}_1,-T/2)
Q^\dagger_{j'} (\boldsymbol{R}, \boldsymbol{y}_2,-T/2) \ket{0},
\nn\\
\\
&& (1) \quad \text{if} \quad \mathcal{M}= T,  \quad \Tensor{\mathcal{M}}^u_{ij}
=\tensor{T}^u_{ij}, \quad u,v=1,2,3\,,
\nn\\
&& (2) \quad \text{if} \quad \mathcal{M}= \Sigma,  \quad \Tensor{\mathcal{M}}^u_{ij}
=\Tensor{\Sigma}^u_{ij}, \quad u,v=1,2,....,6\,,
\nn
\eea
where 
\be
Q(\boldsymbol{R},\boldsymbol{x},t) \equiv 
\phi(\boldsymbol{R},\boldsymbol{x},t) Q(\boldsymbol{x}, t),
\label{Eq:Def_Wilson_String_plus_quark}
\ee
and $\phi(\boldsymbol{R},\boldsymbol{x},t)$ has been defined in Eq.~(\ref{WilsonString}). 
Integrating out the static-quark fields from $I^{uv}_\mathcal{M}$ we obtain 
\begin{equation}
I^{uv}_\mathcal{M} = 
\delta^3(\boldsymbol{x}_1 - \boldsymbol{y}_1)\delta^3 (\boldsymbol{x}_2 - \boldsymbol{y}_2)
\erwzero{(W^\mathcal{M}_{QQ})^{uv}}
\label{Eq:Green_Function_Multiplet_NRQCD_0_0}
\end{equation}
with $W^\mathcal{M}_{QQ}$ diagrammatically represented in Fig.~\ref{fig2mat} 
and explicitly given by 
\begin{equation}
\begin{split}
(W^\mathcal{M}_{QQ})^{uv} \equiv \Pathorder \!\!\!\!\!
\sum_{ijkni'j'k'n'=1}^{3}
\Tensor{\mathcal{M}}^u_{ij} \: \phi_{ii'}(\boldsymbol{R},\boldsymbol{x}_1,T/2)
\phi_{i'k'}(T/2,-T/2,\boldsymbol{x}_1)\phi_{k'k}(\boldsymbol{x}_1,\boldsymbol{R},-T/2) 
\\
{} \times \phi_{jj'}(\boldsymbol{R},\boldsymbol{x}_2,T/2)
\phi_{j'n'}(T/2,-T/2,\boldsymbol{x}_2)\phi_{n'n}(\boldsymbol{x}_2,\boldsymbol{R},-T/2)  
\: \Tensor{\mathcal{M}}^{v}_{kn}.
\end{split}
\end{equation}
\begin{figure}
\begin{center}
\fmfframe(6,6)(6,6){\begin{fmfgraph*}(60,30)
	\fmfforce{(0,h/2)}{v1}
	\fmfforce{(0,h/2)}{t1}	
	\fmfforce{(0,0)}{v2}
	\fmfforce{(w,0)}{v3}
	\fmfforce{(w,h/2)}{v4}
	\fmfforce{(w,h/2)}{t2}
	\fmfforce{(w,h)}{v5}
	\fmfforce{(0,h)}{v6}
	\fmffreeze
	\fmf{plain_arrow}{v1,v2}
	\fmf{plain_arrow}{v2,v3}
	\fmf{plain_arrow}{v3,v4}
	\fmf{plain_arrow}{v1,v6}
	\fmf{plain_arrow}{v6,v5}
	\fmf{plain_arrow}{v5,v4}
	\fmfdot{v1,v4}
	\fmfv{label=$\Tensor{\mathcal{M}}^{v}_{i'j'}$,label.angle=0}{t1}
	\fmfv{label=$\Tensor{\mathcal{M}}^u_{ij}$,label.angle=180}{t2}
	\fmfv{label=$Y$,label.angle=180}{v1}
	\fmfv{label=$y_1$}{v2}
	\fmfv{label=$x_1$}{v3}
	\fmfv{label=$X$,label.angle=0}{v4}
	\fmfv{label=$y_2$}{v6}
	\fmfv{label=$x_2$}{v5}
\end{fmfgraph*}}
\caption{Static Wilson loop with edges $x_1 = ({\boldsymbol{x}}_1,T/2)$, $x_2 = ({\boldsymbol{x}}_2,T/2)$, 
$y_1 = ({\boldsymbol{x}}_1,-T/2)$, $y_2 = ({\boldsymbol{x}}_2,-T/2)$ and insertions of the tensors  
$\Tensor{\mathcal{M}}^u_{ij}$ and $\Tensor{\mathcal{M}}^v_{i'j'}$
in $X = (\boldsymbol{R}, T/2)$ and $Y = (\boldsymbol{R},-T/2)$
respectively.}
\label{fig2mat}
\end{center}
\end{figure}
In the large $T$ limit, the Green functions $I^{uv}_\mathcal{T}$ and
$I^{uv}_{\Sigma}$ are reduced to the antitriplet and sextet propagators of pNRQCD respectively. 
If we neglect subleading loop corrections to the pNRQCD side of the matching, we obtain:
\bea
&&\lim_{T\to\infty} \erwzero{(W^\mathcal{M}_{QQ})^{uv}} =
\lim_{T\to\infty} Z_\mathcal{M}(r) \exp\left(-iV_{\mathcal{M}}^{(0)}(r)\,T\right)
\nn\\
&&\qquad   
\times 
\erwzero{
\sum_{ijkn=1}^{3}\Tensor{\mathcal{M}}^u_{ij} \: \phi_{ik}(T/2,-T/2,\boldsymbol{R})
\phi_{jn}(T/2,-T/2,\boldsymbol{R})\: \Tensor{\mathcal{M}}^{v}_{kn}},
\eea
where $ Z_\mathcal{M}$ is a normalization factor.
At order $\als$ we end up with the well-known result \cite{Flamm:1982jv}:
\begin{align}
V_{T}^{(0)}(r) & = - \frac{2}{3} \frac{\als}{|\boldsymbol{r}|},
\label{VTLO}
\\
V_{\Sigma}^{(0)}(r) & = \frac{1}{3} \frac{\als}{|\boldsymbol{r}|}.
\label{VSLO}
\end{align}
The antitriplet channel is attractive, the sextet one repulsive.

\subsection{Hyperfine Splitting}
In the dynamical situation considered here (case {\bf (A)} of Sec.~\ref{secQQq}), 
a doubly heavy baryon is mainly a bound state of a heavy quark 
(or antiquark) pair in an antitriplet (or triplet) configuration and a light quark (or antiquark). 
Sextet field configurations show up in loops with ultrasoft gluons. 
Their contribution is suppressed either in the multipole expansion 
or in $1/m$ (see Eqs.~(\ref{qqpl01}) and (\ref{qqpl10})). 
The leading-order pNRQCD Lagrangian is shown in Eq.~(\ref{qqLO}).
It does not depend on the spin of the heavy quarks. 
As a consequence, $QQq$ baryons will appear in degenerate multiplets 
of the total spin ${\bf S}_{QQq} = {\bf S}_{QQ} + {\bf S}_l$, 
where ${\bf S}_l$ is the spin of the light degrees of freedom
and ${\bf S}_{QQ}$ of the heavy-quark pair. This symmetry is similar to the
spin symmetry of the HQET. Differently from the HQET, however, the pNRQCD Lagrangian 
depends at leading order on the heavy-quark flavour. This is a consequence 
of the fact that we cannot, in general, neglect the kinetic energy.\footnote{
An exception may be the special case $\lQ \gg mv^2$.} 

We will consider in this section the $S$-wave ground state 
of a doubly heavy baryon made of two identical heavy quarks $Q$ of mass $m_Q$. 
In this case, since an (anti)triplet state is antisymmetric in colour, due 
to the Fermi statistics, the two heavy quarks are allowed only in a spin 1 (symmetric) state.
In the standard notation, the lowest energy states for $QQu$ or $QQd$ 
are called $\Xi_{QQ}$ ($\Xi_{QQ}^{*}$) for spin 
$1/2$ ($3/2$), and for $QQs$, $\Omega_{QQ}$ ($\Omega_{QQ}^{*}$). 
Since the heavy-quark pair spin is fixed, the hyperfine splitting 
may only originate from spin-dependent couplings of the heavy quarks with 
the light one. The leading-order operator (in a $\lQ/m$ expansion) 
is given by Eq.~(\ref{pnrqcdQQSB}).
We will derive a simple formula that relates at leading order in 
the $\lQ/m$ expansion the hyperfine splitting of a $QQq$ doubly heavy baryon 
ground state with the hyperfine splitting of a $\bar{Q}q$ heavy-light meson ground state.
The framework will be that one of pNRQCD, developed in the previous sections.
The calculation will be similar to that one of Ref.~\cite{Savage:di}. 

Let us consider, first, the case of a heavy-light meson $\bar{Q}q$. 
The heavy antiquark may be described by a two-component field 
$Q_c=i \bfsigma_2 Q^*$, where $Q$ is the Pauli spinor that annihilates
the heavy quark. We rename $Q_c^1 = Q_+$ and $Q_c^2 = Q_-$ since 
$Q^\dagger_{\pm} \vert \hbox{0}\rangle =\vert {\bf S}_{\bar{Q}}^z =\pm 1/2\rangle$.
${\bf S}_{\bar{Q}}$ is the spin of the heavy antiquark and ${\bf S}_{\bar{Q}q}$ the total
spin of the meson. The leading-order HQET Lagrangian does not contain spin-interaction terms, 
therefore, states that differ only in the spin quantum numbers are degenerate.
In particular, this happens for the three lowest $S_{\bar{Q}q}=1$ states (${\bf S}^z_{\bar{Q}q} = 1,0,-1$), 
which we denote by $\vert P^*_Q \rangle$, and for the lowest 
$S_{\bar{Q}q}=0$ state, which we denote by $\vert P_Q \rangle$.
An expression for these states that makes explicit their heavy (anti)quark field
content is given in appendix \ref{appC}.
The Hamiltonian responsible for the leading contribution to the hyperfine separation is 
\bea
\delta H_\text{HQET} &=& - c_F^{(Q)} \,\int d^3 R \; \sum_{a=1}^8 Q_c^\dagger({\boldsymbol{R}}) \, 
  \frac{\bfsigma \cdot g{\boldsymbol{B}}^a({\boldsymbol{R}}) \, T_{\bar{3}}^a}{2m_Q} \, Q_c({\boldsymbol{R}}) 
\nn\\
&=&
-\frac{c_F^{(Q)}}{2m_Q} \int d^3 R \, \sum_{a=1}^8 \big [ 
  (Q^\dagger_+ T_{\bar{3}}^a Q_+   - Q^\dagger_{-} T_{\bar{3}}^a Q_{-} )\, g {\boldsymbol{B}}^{3\,a}
+ i (Q^\dagger_{-} T_{\bar{3}}^a Q_{+} - Q^\dagger_+ T_{\bar{3}}^a Q_{-})\,g {\boldsymbol{B}}^{2\,a}
\nn\\
& &\qquad\qquad\quad
+(Q^\dagger_+ T_{\bar{3}}^a Q_{-} + Q^\dagger_{-} T_{\bar{3}}^a Q_{+} ) \,g {\boldsymbol{B}}^{1\,a} 
\big ] ,
\label{shm}
\eea
where, after the last equality, we have dropped the explicit coordinate dependence of the fields. 
From Eq.~(\ref{shm}) and Eqs.~(\ref{spinstat0})-(\ref{spinstatn}) it is straightforward to derive:
\bea
&&\hspace{-8mm}
\langle P^*_Q \vert \delta H_\text{HQET} \vert P^*_Q \rangle - 
\langle P_Q \vert \delta H_\text{HQET} \vert P_Q \rangle  \! = \! 
-2 \frac{c_F^{(Q)}}{m_Q} \!\!
\int \!\! d^3 R \, \langle {\bf S}_l^z =1/2 \vert \sum_{a=1}^8 
g {\boldsymbol{B}}^{3\,a} T_{\bar{3}}^a \vert {\bf S}_l^z =1/2  \rangle.
\label{relqqfinn} 
\eea

In the case of a doubly heavy baryon $QQq$ we proceed in a similar way. 
The triplet field $T$ is a $2\otimes 2$ tensor in spin space, which 
may be decomposed as $2\otimes 2 =  1 \oplus 3$, i.e. 
in a scalar component, $T^{(S)}$, and a vector one, $T^{(V)}$:
\be
T_{ij}({\boldsymbol{r}},{\boldsymbol{R}},t) 
=  \left(\frac{i\bfsigma_2}{\sqrt{2}} \right)_{ij} T^{(S)}({\boldsymbol{r}},{\boldsymbol{R}},t) 
+ \sum_{k=1}^3 \left( \frac{i\bfsigma_k\bfsigma_2}{\sqrt{2}} \right)_{ij}
T^{(V)\,k}({\boldsymbol{r}},{\boldsymbol{R}},t), \qquad i,j = 1,2.
\label{spindecnQQ}
\ee
The indices $ij$ refer to the spin space. Note that the matrices $(i\bfsigma_2/\sqrt{2})_{ij}$ and 
$(i\bfsigma_k\bfsigma_2/\sqrt{2})_{ij}$ are respectively antisymmetric and symmetric in
$ij$. It is convenient to rewrite the fields $T^{(V)\,k}$ as 
\be 
T_0 = T^{(V)\,3} \qquad \text{and} \qquad
T_{\pm} = \frac{\mp T^{(V)\,1}+iT^{(V)\,2}}{\sqrt{2}},
\label{defTpm0}
\ee
since $T^\dagger_{0} \vert \hbox{0}\rangle =\vert {\bf S}_{QQ}^z =0\rangle$ and 
$T^\dagger_{\pm} \vert \hbox{0}\rangle =\vert {\bf S}_{QQ}^z =\pm 1 \rangle$.
As we argued above, the leading-order pNRQCD Lagrangian describing doubly heavy baryons 
does not contain spin-interaction terms, 
therefore, states that differ only in the spin quantum numbers are degenerate.
In particular, this happens for the four lowest $S_{QQq}=3/2$ states
(${\bf S}^z_{QQq} = \pm 3/2, \pm 1/2$), which we denote by $\vert \Xi^*_{QQ} \rangle$, and for
the two lowest $S_{QQq}=1/2$ states (${\bf S}^z_{QQq} = \pm 1/2$), which we denote by $\vert \Xi_{QQ} \rangle$.
An explicit expression of these states in terms of heavy (anti)triplet fields is given in appendix \ref{appC}.
The Hamiltonian responsible for the leading contribution to the hyperfine separation is 
\bea
\delta H_\text{pNRQCD} 
&=& - \frac{c_F^{(Q)}}{2 m_Q} \int d^3 R  \int d^3 r \;  V_{T\bfsigma\cdot{\bf
    B}T}^{(1,0)}(r) 
\sum_{a=1}^8
T^\dagger ({\boldsymbol{r}},{\boldsymbol{R}})\;
\frac{\bfsigma^{(1)} + \bfsigma^{(2)}}{2} \cdot g{\boldsymbol{B}}^a({\boldsymbol{R}}) T^a_{\bar{3}} 
\; T ({\boldsymbol{r}},{\boldsymbol{R}}) 
\nn\\
&=& - \frac{c_F^{(Q)}}{2 m_Q} \int d^3 R  \int d^3 r \;  V_{T\bfsigma\cdot{\bf
    B}T}^{(1,0)}(r) 
\sum_{a=1}^8\sum_{ikj=1}^3
T^{(V)\,i\,\dagger}\; i \epsilon_{ikj}  \;  g {\boldsymbol{B}}^{k\,a} T_{\bar{3}}^a \; T^{(V)\,j} 
\nn
\\
&=&
-\frac{c_F^{(Q)}}{ 2m_Q} \int d^3 R \int d^3 r \; V_{T\bfsigma\cdot{\boldsymbol{B}}T}^{(1,0)}(r)\sum_{a=1}^8
\bigg[ (T_{+}^\dagger T_{\bar{3}}^a T_{+} - T_{-}^\dagger T_{\bar{3}}^a T_{-})
  \, g {\boldsymbol{B}}^{3\, a} 
\nn
\\
& & \qquad\quad  
+\frac{i}{ \sqrt{2}} (-T_{+}^\dagger T_{\bar{3}}^a T_{0} +T_{0}^\dagger T_{\bar{3}}^a T_{+} 
+T_{-}^\dagger T_{\bar{3}}^a T_{0} - T_{0}^\dagger T_{\bar{3}}^a T_{-}) \, g {\boldsymbol{B}}^{2\, a} 
\nn
\\
& & \qquad\quad  
+  \frac{1}{ \sqrt{2}}  
(T_{+}^\dagger T_{\bar{3}}^a T_{0} + T_{0}^\dagger T_{\bar{3}}^a T_{+} 
+T_{-}^\dagger T_{\bar{3}}^a T_{0} + T_{0}^\dagger T_{\bar{3}}^a T_{-}) \, g
{\boldsymbol{B}}^{1\, a} \bigg],
\label{n2qqqspin}
\eea
where the second equality follows from Eq.~(\ref{spindecnQQ}) and the third
one from Eq.~(\ref{defTpm0}).
From Eq.~(\ref{n2qqqspin}) and Eqs.~(\ref{spinstatnqq0})-(\ref{spinstatnqq}) 
it is straightforward to derive:
\bea
&&\hspace{-8mm}
\langle \Xi^*_{QQ} \vert \delta H_\text{pNRQCD}  \vert \Xi^*_{QQ} \rangle -
\langle \Xi_{QQ} ; \vert \delta H_\text{pNRQCD} \vert \Xi_{QQ} \rangle 
=
\nn\\
&&
- 3 \frac{c_F^{(Q)} }{ 2m_Q} 
\int \! d^3 R \, \langle {\bf S}_l^z =1/2 \vert 
\sum_{a=1}^8 g {\boldsymbol{B}}^{3\,a} T_{\bar{3}}^a \vert {\bf S}_l^z =1/2  \rangle 
\int \! d^3 r \, \varphi^*_{QQ}({\boldsymbol{r}})\, V_{T\bfsigma\cdot{\bf
    B}T}^{(1,0)}(r) \, \varphi_{QQ}({\boldsymbol{r}}),
\label{hfsQQq}
\eea
where $ \varphi_{QQ}$ is the ground-state eigenfunction of
$-\bfnabla^2_r/(2m_r) + V_T^{(0)}$.
At NLO $ V_{T\bfsigma\cdot{\boldsymbol{B}}T}^{(1,0)} =1$, therefore 
$\displaystyle \int d^3 r \; \varphi^*_{QQ}({\boldsymbol{r}})\, V_{T\bfsigma\cdot{\bf
    B}T}^{(1,0)}(r) \, \varphi_{QQ}({\boldsymbol{r}}) = 1 + {\cal O}(\als^2)$.
This result crucially depends on the fact that we have 
multipole expanded the gluon fields. As a consequence, ${\boldsymbol{B}}$ does not
depend on ${\boldsymbol{r}}$ and the magnetic dipole transition term (differently from 
the electric one $\propto {\boldsymbol{r}}\cdot g{\boldsymbol{E}}$) 
does not exhibit any explicit dependence on ${\boldsymbol{r}}$.
Comparing Eq.~(\ref{relqqfinn}) with Eq.~(\ref{hfsQQq}) we obtain ($M_\Xi$ and
$M_P$ are the baryon and meson masses respectively) 
\be
M_{\Xi^*_{QQ}}-M_{\Xi_{QQ}} = \frac{3\, m_{Q'}}{4\, m_Q}\frac{c_F^{(Q)}}{c_F^{(Q')}} 
\left(M_{P^*_{Q'}}-M_{P_{Q'}}\right)
\left[1+{\cal O}\left(\als^2, \frac{\lQ}{m_Q} , \frac{\lQ}{m_{Q'}} \right)\right].
\label{relff}
\ee
Up to a factor $1/2$, the formula is the one derived in \cite{Savage:di}.
In the previous section, the origin of the discrepancy has been traced back 
to a missing colour normalization factor in the spin antitriplet interaction 
term (\ref{pnrqcdQQSB}) (and (\ref{n2qqqspin})).
On the other hand, the relation 
$\displaystyle M_{\Xi^*_{QQ}}-M_{\Xi_{QQ}} = \frac{3\, m_{Q'}}{4\, m_Q}
\left(M_{P^*_{Q'}}-M_{P_{Q'}}\right)$ has been derived since long in  
non-relativistic potential models. Surprisingly the discrepancy between 
this formula and the formula in \cite{Savage:di} has to the best of our knowledge 
never been noticed before in the literature.\footnote{From private communications 
we know, however, that at least Tom Mehen and the authors of \cite{Lewis:2001iz}
were aware of it.}
Even more surprisingly some of the literature has explicitly claimed agreement 
between the potential model prediction and the formula in \cite{Savage:di}!

From \cite{Eidelman:2004wy} we read that $M_{D^*}-M_{D} = 142.12\pm0.07$ MeV 
and $M_{B^*}-M_{B} = 45.78\pm0.35$ MeV. Both data may be used 
to obtain $M_{\Xi^*_{cc}}-M_{\Xi_{cc}}$ and $M_{\Xi^*_{bb}}-M_{\Xi_{bb}}$
from Eq.~(\ref{relff}). If $Q \neq Q'$, we use $c_F^{(Q)}$ at NLL accuracy calculated 
in \cite{Amoros:1997rx}, and $m_b = M_{\Upsilon(1S)}/2$ and  
$m_c =M_{J/\psi}/2$. For $M_{\Xi^*_{cc}}-M_{\Xi_{cc}}$ we obtain 
about 107 MeV from the $D$ data and about 133 MeV from the $B$ data. 
Taking the average and estimating $\lQ/m_c \sim \als^2(m_c\als) \approx 0.3$, our
result is:
\be
M_{\Xi^*_{cc}}-M_{\Xi_{cc}} = 120 \pm 40 \;\; {\rm MeV}.
\label{Xicc}
\ee
Similarly for $M_{\Xi^*_{bb}}-M_{\Xi_{bb}}$ we obtain 
about 27 MeV from the $D$ data and about 34 MeV from the $B$ data. 
Taking only the estimate based on the $B$ data, because affected by the smaller 
uncertainty $\lQ/m_b \sim \als^2(m_b\als) \approx 0.1$, our result is 
\be
M_{\Xi^*_{bb}}-M_{\Xi_{bb}} = 34 \pm 4 \;\; {\rm MeV}.
\label{Xibb}
\ee
These results compare well with the quenched QCD lattice simulation of 
\cite{Flynn:2003vz}, whose result is $M_{\Xi^*_{cc}}-M_{\Xi_{cc}}= 89 \pm 15$ MeV, 
and of \cite{Lewis:2001iz}, whose result is  $M_{\Xi^*_{cc}}-M_{\Xi_{cc}}= 80
\pm 10^{+3}_{-7}$ MeV, and with the quenched NRQCD lattice simulations of
\cite{AliKhan:1999yb} and \cite{Mathur:2002ce}, whose results   
for $bbq$ baryons are  $M_{\Xi^*_{bb}}-M_{\Xi_{bb}}= 20 \pm 6^{+2}_{-3}$ MeV and 
$M_{\Xi^*_{bb}}-M_{\Xi_{bb}}= 20 \pm 6^{+3}_{-4}$ MeV respectively.
The figures of \cite{Lewis:2001iz} and \cite{Mathur:2002ce} 
refer to the lattice calculations at largest $\beta$.

\section{pNRQCD for $QQQ$  Baryons}
\label{secQQQ}
In this section we consider baryons formed by three heavy quarks of which at
least two with the same mass $m_1=m_2\equiv m$. Baryons of this type may be
composed by $bbb$, $bbc$, $bcc$ or $ccc$ quarks.
We define
\bea
&&m_R = 2\, m+m_3\,, \qquad\qquad\quad m_\rho = \frac{m}{2}\,, \qquad\> m_\lambda= \frac{2\, m \,m_3}{m_R}\,,
\\
&&{\boldsymbol{R}} = \frac{m({\boldsymbol{x}}_1+{\boldsymbol{x}}_2) + m_3 {\boldsymbol{x}}_3}{m_R}\,,
\quad \boldsymbol{\rho} = \boldsymbol{x}_1 - \boldsymbol{x}_2\,,
\quad \boldsymbol{\lambda} = \frac{\boldsymbol{x}_1 + \boldsymbol{x}_2}{2} -  \boldsymbol{x}_3\,.
\eea
There are, in principle, several physical scales that may play an important role in the dynamics: 
the masses $m_R$, $m_\rho$  and $m_\lambda$, which we assume to be of the same order, the
typical relative three momenta of the heavy quarks, the typical kinetic energies 
and the scale of non-perturbative physics $\lQ$.
In the following, we will keep the discussion as simple as possible by 
not exploiting any possible hierarchy among the relative momenta 
and the kinetic energies. We will assume that the typical relative momenta
of the heavy quarks, generically denoted by $mv$, are all much smaller 
than the heavy-quark masses and much larger than the kinetic energies, 
generically denoted by $mv^2$. We may distinguish two situations.

{\bf (A)} The typical relative momenta of the heavy quarks are much larger 
than $\lQ$. We call this situation weakly coupled.

{\bf (B)} The typical relative momenta of the heavy quarks are of the order of 
$\lQ$. We call this situation strongly coupled.

\subsection{${\rm pNRQCD}$ for weakly-coupled $QQQ$ baryons}

\subsubsection{Lagrangian and Degrees of Freedom}
If we assume that the typical distances  $\rho$ and $\lambda$ in the baryon,
which are of order $1/(mv)$, are much smaller than $1/\lQ$, then  
gluons of momentum or energy of order $mv$ may be integrated out 
from NRQCD order by order in $\als$.
The resulting EFT has light quarks, gluons of energy and momentum lower than $mv$
(ultrasoft gluons), and  heavy quarks as degrees of freedom.
Gluons appearing in vertices involving heavy-quark fields are multipole
expanded in $\boldsymbol{\rho}$ and $\boldsymbol{\lambda}$ to ensure that they are ultrasoft.
This corresponds to expanding in $\rho\lQ \sim \lQ/(mv)$ and $\lambda\lQ \sim
\lQ/(mv)$, or $\rho\,(mv^2) \sim v$ and  $\lambda\,(mv^2) \sim v$.
Like in the case of doubly heavy baryons  
the number of allowed operators is consistently reduced if we
choose to have a manifestly gauge-invariant Lagrangian. 
This may be done by projecting the Lagrangian on the heavy-heavy-heavy sector 
of the Fock space, by splitting the heavy-heavy-heavy fields into a 
singlet, two octet and a decuplet component,
\bea 
&& \hspace{-10mm}
Q_{1\,i}({\boldsymbol{x}}_1,t) 
Q_{2\,j}({\boldsymbol{x}}_2,t) 
Q_{3\,k}({\boldsymbol{x}}_3,t) 
\sim
S(\boldsymbol{\rho},\boldsymbol{\lambda},{\boldsymbol{R}},t) \, \tensor{S}_{ijk} 
+ \sum_{a=1}^8 O^{\text{A}\,a}(\boldsymbol{\rho},\boldsymbol{\lambda},{\boldsymbol{R}},t) 
\, \tensor{O}^{\text{A}\,a}_{ijk}
\nn\\
&& \qquad\qquad
+ \sum_{a=1}^8 O^{\text{S}\,a}(\boldsymbol{\rho},\boldsymbol{\lambda},{\boldsymbol{R}},t) 
\, \tensor{O}^{\text{S}\,a}_{ijk}
+ \sum_{\delta=1}^{10} \Delta^{\delta}(\boldsymbol{\rho},\boldsymbol{\lambda},{\boldsymbol{R}},t) 
\, \Tensor{\Delta}^{\delta}_{ijk}, \quad
i,j=1,2,3\,, 
\label{QQQTS}
\eea
and by building the Lagrangian from these operators.
The tensors $\tensor{S}_{ijk}$, $\tensor{O}^{\text{A}\,a}_{ijk}$, 
$\tensor{O}^{\text{S}\,a}_{ijk}$  and $\Tensor{\Delta}^{\delta}_{ijk}$ 
are defined in appendix \ref{App:Multiplet_TensorsQQQ}.
$\tensor{S}_{ijk}$ and $\Tensor{\Delta}^{\delta}_{ijk}$ are real and
respectively totally antisymmetric and symmetric. We chose 
$\tensor{O}^{\text{A}\,a}_{ijk}$ and $\tensor{O}^{\text{S}\,a}_{ijk}$ to be 
respectively antisymmetric and symmetric in the first two indices.
The Lagrangian is constrained to satisfy all the symmetries 
of QCD. In particular, in the case $m_1=m_2$ that we consider here, 
it must be invariant under the exchange of the heavy quarks labeled 1 and 2.
Under such transformation, ${\bfrho}$ goes into $-{\bfrho}$ and ${\bflambda}$ goes into ${\bflambda}$. 
The gluon fields are even, because multipole 
expanded around the centre-of-mass of the heavy-heavy-heavy system.
For what concerns the heavy-quark fields, from Eq.~(\ref{QQQTS}) it follows that 
$S$ and $O^{\text{A}\,a}$ are even, because $\tensor{S}_{ijk}$ and 
$ \tensor{O}^{\text{A}\,a}_{ijk}$ are odd under exchange $i
\leftrightarrow j$, and $O^{\text{S}\,a}$ and $\Delta^{\delta}$
are odd, because $\tensor{O}^{\text{S}\,a}_{ijk}$ and
$\Tensor{\Delta}^{\delta}_{ijk}$ are even under exchange $i
\leftrightarrow j$. It is also useful to consider the combination 
of the above transformation with parity, $(1\leftrightarrow 2) \times P$, 
which is also a symmetry of the Lagrangian.
Under this transformation ${\bfrho}$ goes into ${\bfrho}$ 
and ${\bflambda}$ goes into $-{\bflambda}$. The gluon fields 
transform like $A_\mu(t,{\boldsymbol{R}}) \rightarrow A^\mu(t,-{\boldsymbol{R}})$, 
which means that, up to reflection of the internal spatial coordinates, 
chromoelectric fields are odd and chromomagnetic fields are even.
Up to reflection of the internal spatial coordinates, the 
heavy-quark fields transform like in the case of the $(1\leftrightarrow 2)$ exchange.

The resulting Lagrangian $\mathcal{L}_\text{pNRQCD} 
= \mathcal{L}_\text{pNRQCD}({\boldsymbol{R}},t)$ at ${\cal O}(\lambda,\rho)$ in the multipole expansion 
(we also display at ${\cal O}(1/m)$ the kinetic energy terms) is 
\begin{equation}
\mathcal{L}_\text{pNRQCD} = 
\mathcal{L}_\text{gluon} 
+ \mathcal{L}_\text{light} 
+ \mathcal{L}^{(0,0)}_\text{pNRQCD} 
+ \mathcal{L}^{(0,1)}_\text{pNRQCD} 
+ \mathcal{L}^{(1,0)}_\text{pNRQCD}, 
\label{pl}
\end{equation}
with $\mathcal{L}_\text{gluon}$ and $\mathcal{L}_\text{light}$ defined in Eqs.~(\ref{Lgluon}) 
and (\ref{Llight}) respectively and 
\begin{align}
&\mathcal{L}^{(0,0)}_\text{pNRQCD} = \int d^3\rho\,d^3\lambda \;
\: S^\dagger \left[ i\partial_0 - V^{(0)}_S \right] S 
+ O^{\text{A}\dagger} \left[ iD_0 - V^{(0)}_{O^\text{A}} \right] O^\text{A} 
\notag\\
& \hspace{3mm}
+ O^{\text{S}\dagger} \left[ iD_0 - V^{(0)}_{O^\text{S}} \right] O^\text{S} 
+ \Delta^\dagger \left[ iD_0 - V^{(0)}_{\Delta} \right] \Delta \,,
\label{pl0}
\end{align}
\begin{align}
&\mathcal{L}^{(0,1)}_\text{pNRQCD}  = 
\int d^3\rho\,d^3\lambda  \; 
V^{(0,1)}_{S \bfrho\cdot{\boldsymbol{E}} O^\text{S}} 
\sum_{a=1}^8 \frac{1}{2\sqrt{2}}
\left[ S^\dagger \boldsymbol{\rho}\cdot g\boldsymbol{E}^a\,O^{\text{S}\,a} 
+ O^{\text{S}\,a\dagger} \boldsymbol{\rho}\cdot g\boldsymbol{E}^a \,S \right] 
\notag\\
& \hspace{3mm}
- V^{(0,1)}_{O^\text{A} \bfrho\cdot{\boldsymbol{E}} O^\text{S}}
\sum_{abc=1}^8
\left(\frac{if^{abc}+3d^{abc}}{4\sqrt{3}}\right)
\left[ O^{\text{A}\,a\dagger} \boldsymbol{\rho}\cdot g\boldsymbol{E}^b\,O^{\text{S}\,c} 
+ O^{\text{S}\,a\dagger} \boldsymbol{\rho}\cdot g\boldsymbol{E}^b\,O^{\text{A}\,c} \right] 
\notag\\
& \hspace{3mm}
+ V^{(0,1)}_{O^\text{A}\bfrho\cdot{\boldsymbol{E}}\Delta}
\sum_{ab=1}^8\sum_{\delta=1}^{10}
\left[ 
\left(\sum_{ii'jj'k=1}^3\epsilon_{ijk}T^a_{ii'}T^b_{jj'}\Tensor{\Delta}^\delta_{i'j'k}\right)\;
O^{\text{A}\,a\dagger} \boldsymbol{\rho}\cdot g\boldsymbol{E}^b\,\Delta^\delta 
\right.
\notag\\
&\qquad \qquad \qquad \qquad \qquad \qquad \qquad {}
\left.
- \left(\sum_{ii'jj'k=1}^3\Tensor{\Delta}^\delta_{ijk}T^b_{ii'}T^a_{jj'}\epsilon_{i'j'k}\right)\;
\Delta^{\delta\dagger} \boldsymbol{\rho}\cdot g\boldsymbol{E}^b\,O^{\text{A}\,a} \right]
\notag \\
&\hspace{3mm}
- V^{(0,1)}_{S \bflambda\cdot{\boldsymbol{E}} O^\text{A}}
\sum_{a=1}^8 \frac{1}{\sqrt{6}}
\left[ S^\dagger \boldsymbol{\lambda}\cdot g\boldsymbol{E}^a \,O^{\text{A}\,a} 
+ O^{\text{A}\,a\dagger} \boldsymbol{\lambda}\cdot g\boldsymbol{E}^a\,S \right] 
\notag\\
& \hspace{3mm}
- V^{(0,1)}_{O^{\text{A}} \bflambda\cdot{\boldsymbol{E}} O^\text{A}}
\sum_{abc=1}^8 \left( if^{abc}\frac{2m-m_3}{2m_R} + \frac{d^{abc}}{2} \right) \;
O^{\text{A}\,a\dagger} \boldsymbol{\lambda}\cdot g\boldsymbol{E}^b \,O^{\text{A}\,c} 
\notag\\
&\hspace{3mm}
+ V^{(0,1)}_{O^\text{S} \bflambda\cdot{\boldsymbol{E}} O^\text{S}} 
\sum_{abc=1}^8 \left( if^{abc}\frac{5m_3-2m}{6m_R} + \frac{d^{abc}}{2} \right) 
O^{\text{S}\,a\dagger} \boldsymbol{\lambda}\cdot g\boldsymbol{E}^b \,O^{\text{S}\,c} 
\notag\\
& \hspace{3mm}
+ V^{(0,1)}_{O^\text{S} \bflambda\cdot{\boldsymbol{E}} \Delta}
\sum_{ab=1}^8\sum_{\delta=1}^{10}
\left[ \left(\frac{2}{\sqrt{3}}\sum_{ii'jj'k=1}^3\epsilon_{ijk}T^a_{ii'}T^b_{jj'}\Tensor{\Delta}^\delta_{i'j'k}\right)\;
O^{\text{S}\,a\dagger} \boldsymbol{\lambda}\cdot\boldsymbol{E}^b\,\Delta^\delta 
\right.
\notag\\
&\qquad \qquad \qquad \qquad \qquad \qquad \qquad  {}
\left.
- \left(\frac{2}{\sqrt{3}}\sum_{ii'jj'k=1}^3\Tensor{\Delta}^\delta_{ijk}T^b_{ii'}T^a_{jj'}\epsilon_{i'j'k}\right)\;
\Delta^{\delta\dagger} \boldsymbol{\lambda}\cdot g\boldsymbol{E}^b \,O^{\text{S}\,a} \right] 
\notag\\
& \hspace{3mm}
- V^{(0,1)}_{\Delta  \bflambda\cdot{\boldsymbol{E}} \Delta} 
\frac{2m-2m_3}{m_R}
\sum_{a=1}^8
\sum_{\delta\delta'=1}^{10} 
\left(\sum_{ii'jk=1}^3 \Tensor{\Delta}^\delta_{ijk}T^a_{ii'}\Tensor{\Delta}^{\delta'}_{i'jk}\right)\;
\Delta^{\delta\dagger} \boldsymbol{\lambda}\cdot g\boldsymbol{E}^a\,\Delta^{\delta'} \,,
\label{pl01}
\\
& \nn\\
&\mathcal{L}^{(1,0)}_\text{pNRQCD} = \int d^3\rho\,d^3\lambda \;
\: S^\dagger 
\left[ \frac{\boldsymbol{\nabla}^2_R}{ 2m_R} 
+ \frac{\boldsymbol{\nabla}^2_\rho }{ 2m_\rho} 
+ \frac{\boldsymbol{\nabla}^2_\lambda }{ 2m_\lambda} 
\right ] S 
+ O^{\text{A}\dagger} 
\left[ \frac{\boldsymbol{D}^2_R}{2m_R} 
+ \frac{\boldsymbol{\nabla}^2_\rho}{2m_\rho} 
+ \frac{\boldsymbol{\nabla}^2_\lambda}{2m_\lambda} 
\right] O^\text{A} 
\notag\\
& \hspace{3mm}
+ O^{\text{S}\dagger} 
\left[ \frac{\boldsymbol{D}^2_R}{2m_R} 
+ \frac{\boldsymbol{\nabla}^2_\rho}{2m_\rho} 
+ \frac{\boldsymbol{\nabla}^2_\lambda}{2m_\lambda} 
\right] O^\text{S} 
+ \Delta^\dagger 
\left[ \frac{\boldsymbol{D}^2_R}{2m_R} 
+ \frac{\boldsymbol{\nabla}^2_\rho}{2m_\rho} 
+ \frac{\boldsymbol{\nabla}^2_\lambda}{2m_\lambda} 
\right] \Delta 
\notag\\
& \hspace{3mm}
+ \cdots \quad,
\label{pl10}
\end{align}
where the gauge fields in the covariant derivatives acting on the octets,  
$O^\text{A}$ $=$ $(O^{\text{A}\,1},$ $O^{\text{A}\,2},\dots,$ $O^{\text{A}\,8})$
and $O^\text{S}$ $=$ $(O^{\text{S}\,1},$ $O^{\text{S}\,2},\dots,$ $O^{\text{S}\,8})$, 
and the decuplet, $\Delta$ $=$ $(\Delta^{1},$ $\Delta^{2},\dots,$ $\Delta^{10})$, 
are understood in the octet and decuplet representation respectively.
The dots in the last line of Eq.~(\ref{pl10}) stand for terms that appear at
orders higher than tree level and other $1/m$ terms, similar to those 
discussed for the doubly heavy baryon case. 
These terms are suppressed in the power counting with respect
to the kinetic energy and the terms shown in Eqs.~(\ref{pl0}) and (\ref{pl01}).

The functions $V$ are the Wilson coefficients of pNRQCD. They encode the contributions 
coming from gluons of energy or momentum of order $mv$.
They are non-analytic functions of $\bfrho$ and $\bflambda$. As we will
discuss in the next section,  at tree level we have
\bea 
&& V^{(0,1)}_{S \bfrho\cdot{\boldsymbol{E}} O^\text{S}} = 
V^{(0,1)}_{O^\text{A} \bfrho\cdot{\boldsymbol{E}} O^\text{S}} = 
V^{(0,1)}_{O^\text{A}\bfrho\cdot{\boldsymbol{E}}\Delta} =1\,,
\nn\\
&& 
V^{(0,1)}_{S \bflambda\cdot{\boldsymbol{E}} O^\text{A}} =
V^{(0,1)}_{O^\text{A} \bflambda\cdot{\boldsymbol{E}} O^\text{A}} =
V^{(0,1)}_{O^\text{S} \bflambda\cdot{\boldsymbol{E}} O^\text{S}} =
V^{(0,1)}_{O^\text{S} \bflambda\cdot{\boldsymbol{E}} \Delta} =
V^{(0,1)}_{\Delta  \bflambda\cdot{\boldsymbol{E}} \Delta} =1\,, 
\label{QQQtree}
\eea
while $V^{(0)}_S$, $V^{(0)}_{O^{\text{A}}}$, $V^{(0)}_{O^{\text{S}}}$  and $V^{(0)}_{\Delta}$ 
get the first non-vanishing contribution at order $\als$. 
The coefficients in front of the operators $\boldsymbol{D}_R^2$, $\bfnabla^2_\rho$ and $\bfnabla^2_\lambda$
are equal to 1, due to Poincar\'e invariance or dynamical considerations 
similar to those developed in \cite{Brambilla:2003nt}.

The power counting of the Lagrangian (\ref{pl}) in the centre-of-mass frame  goes as follows: 
$\bfnabla_\lambda, \bfnabla_\rho \sim mv$, $\bfrho, \bflambda \sim 1/(mv)$, ${\boldsymbol{D}}_{R} \sim \lQ, mv^2$, 
$V^{(0)}_{S, O^{\text{A,S}},\Delta} \sim mv^2$ and ${\boldsymbol{E}},{\boldsymbol{B}} \sim \lQ^2, (mv^2)^2$.
The pNRQCD Lagrangian at leading order reads:
\bea
&& \mathcal{L}^\text{LO}_\text{pNRQCD} = 
\int d^3\rho\,d^3\lambda \;\left\{
\: S^\dagger \left[ i\partial_0 
+ \frac{\boldsymbol{\nabla}^2_\rho }{ 2m_\rho} 
+ \frac{\boldsymbol{\nabla}^2_\lambda }{ 2m_\lambda} 
- V^{(0)}_S \right] S \right.
\nn\\
&& \qquad  + O^{\text{A}\dagger} \left[ iD_0 
+ \frac{\boldsymbol{\nabla}^2_\rho }{ 2m_\rho} 
+ \frac{\boldsymbol{\nabla}^2_\lambda }{ 2m_\lambda} 
- V^{(0)}_{O^\text{A}} \right] O^\text{A} 
+ O^{\text{S}\dagger} \left[ iD_0 
+ \frac{\boldsymbol{\nabla}^2_\rho }{ 2m_\rho} 
+ \frac{\boldsymbol{\nabla}^2_\lambda }{ 2m_\lambda} 
- V^{(0)}_{O^\text{S}} \right] O^\text{S} 
\nn\\
&& \qquad {} + \Delta^\dagger \left.\left[ iD_0 
+ \frac{\boldsymbol{\nabla}^2_\rho }{ 2m_\rho} 
+ \frac{\boldsymbol{\nabla}^2_\lambda }{ 2m_\lambda} 
- V^{(0)}_{\Delta} \right] \Delta \right\}
-\frac{1}{4}  \sum_{a=1}^8 F^a_{\mu\nu}F^{a\mu\nu}
+ \sum_{f=1}^3 \bar{q}_f \, i \dsl \,  q_f.
\label{qqqLO}
\eea

\subsubsection{Matching}
The matching from NRQCD to pNRQCD is performed by 
calculating Green functions in the two theories and imposing that 
they are equal order by order in the inverse of the mass and in the multipole expansion.
Since we are working here in the situation where the typical momentum transfer between 
the heavy quarks is larger than $\lQ$, we can in addition
perform the matching order by order in $\als$.
The procedure is analogous to that one discussed previously 
for the doubly heavy baryon case, which we follow closely.

The matching at tree level may be performed by 
projecting the NRQCD Hamiltonian on the three-quark Fock space spanned by 
\be
\int d^3x_1 \, d^3x_2 \, d^3x_3 \, \sum_{ijk=1}^3
\Phi_{Q_1Q_2Q_3}^{ijk}(\boldsymbol{x}_1, \boldsymbol{x}_2, \boldsymbol{x}_3)
Q^{i\,\dagger}_1(\boldsymbol{x}_1)
Q^{j\,\dagger}_2(\boldsymbol{x}_2)
Q^{k\,\dagger}_3(\boldsymbol{x}_3) \ket{0},
\label{Eq:Arbitrary_Three_Particle_State}
\ee
where $\Phi_{Q_1Q_2Q_3}(\boldsymbol{x}_1,\boldsymbol{x}_2,\boldsymbol{x}_3)$ 
is a $3 \otimes 3 \otimes 3$ tensor in colour space and a $2 \otimes 2 \otimes
2$ tensor in spin space. After projection, all gluon fields are multipole 
expanded in $\bfrho$ and $\bflambda$. In order to make gauge invariance
explicit at the Lagrangian level, we decompose the three quark fields into a 
field $S(\boldsymbol{\rho}, \boldsymbol{\lambda}, \boldsymbol{R}, t)$, 
which transforms like a colour singlet, two fields 
$O^\text{A}(\boldsymbol{\rho}, \boldsymbol{\lambda}, \boldsymbol{R}, t)$ and 
$O^\text{S}(\boldsymbol{\rho}, \boldsymbol{\lambda}, \boldsymbol{R}, t)$, which transform like octets,  
and  a field $\Delta(\boldsymbol{\rho}, \boldsymbol{\lambda}, \boldsymbol{R}, t)$, which transforms
like a decuplet:
\bea
\Phi_{Q_1Q_2Q_3}^{ijk} (\boldsymbol{x}_1, \boldsymbol{x}_2,
\boldsymbol{x}_3,t) &=& \sum_{i'j'k'=1}^{3}
\phi_{ii'} (\boldsymbol{x}_1, \boldsymbol{R}, t) \,
\phi_{jj'} (\boldsymbol{x}_2, \boldsymbol{R}; t) \,
\phi_{kk'} (\boldsymbol{x}_3, \boldsymbol{R}; t) \,
\nn\\
&& \times 
\left( S(\boldsymbol{\rho}, \boldsymbol{\lambda}, \boldsymbol{R}, t)\, \tensor{S}_{i'j'k'}
+ \sum_{a=1}^8 O^{\text{A}\,a}(\boldsymbol{\rho}, \boldsymbol{\lambda}, \boldsymbol{R}, t)
\,  \tensor{O}^{\text{A}\,a}_{i'j'k'}
\right.
\nn\\
&& \hspace{3mm}
\left.
+ \sum_{a=1}^8 O^{\text{S}\,a}(\boldsymbol{\rho}, \boldsymbol{\lambda}, \boldsymbol{R}, t)
\,  \tensor{O}^{\text{S}\,a}_{i'j'k'}
+ \sum_{\delta=1}^{10} \Delta^{\delta}(\boldsymbol{\rho}, \boldsymbol{\lambda}, \boldsymbol{R}, t)
\,  \Tensor{\Delta}^{\delta}_{i'j'k'}
\right),
\label{stateQQQ}
\eea
where $ \tensor{S}_{ijk}$, $\tensor{O}^{\text{A}\,a}_{ijk}$, $\tensor{O}^{\text{S}\,a}_{ijk}$
and $\Tensor{\Delta}^{\delta}_{ijk}$ have been defined in appendix
\ref{App:Multiplet_TensorsQQQ} and the Wilson string $\phi_{ij}$ in Eq.~(\ref{WilsonString}).
After projecting on (\ref{stateQQQ}) the Lagrangian (\ref{pl0})-(\ref{pl10})
with the matching conditions (\ref{QQQtree}) follows.

The perturbative matching of the static potentials 
$V^{(0)}_S$, $V^{(0)}_{O^\text{A}}$, $V^{(0)}_{O^\text{S}}$ and $V^{(0)}_{\Delta}$ 
goes as follows. In NRQCD we compute static Green functions, whose initial and
final states overlap with the singlet, octet and decuplet fields in pNRQCD.
A possible choice, working in a non-gauge invariant framework, is
\bea
&&\hspace{-5mm} I^{uv}_\mathcal{M} \equiv \!\!\!
\sum_{ijki'j'k'=1}^{3}
\bra{0} \Tensor{\mathcal{M}}^u_{ijk} \, 
Q_i(\boldsymbol{R},\boldsymbol{x}_1,T/2) 
Q_j(\boldsymbol{R}, \boldsymbol{x}_2, T/2) 
Q_k(\boldsymbol{R}, \boldsymbol{x}_3, T/2) 
\nn\\
&&\hspace{20mm}
\times \Tensor{\mathcal{M}}^{v\,*}_{i'j'k'} \,
Q^\dagger_{i'} (\boldsymbol{R}, \boldsymbol{y}_1,-T/2)
Q^\dagger_{j'} (\boldsymbol{R}, \boldsymbol{y}_2,-T/2) 
Q^\dagger_{k'} (\boldsymbol{R}, \boldsymbol{y}_3,-T/2) 
\ket{0},
\\
\nn
\eea
\bea
&& \qquad 
(1) \quad \text{if} \quad \mathcal{M}=S,  \quad I^{uv}_\mathcal{M} = I_S, 
\quad \Tensor{\mathcal{M}}^u_{ijk} =\tensor{S}_{ijk}, 
\nn\\
&& \qquad 
(2) \quad \text{if} \quad \mathcal{M}= O^\text{A},  \quad I^{uv}_\mathcal{M}
=I^{uv}_{O^\text{A}},  \quad \Tensor{\mathcal{M}}^u_{ijk}
=\tensor{O}^{\text{A}\,u}_{ijk}, \qquad 
u,v=1,2, ..., 8\,,
\nn\\
&& \qquad 
(3) \quad \text{if} \quad \mathcal{M}= O^\text{S}, \quad I^{uv}_\mathcal{M}
=I^{uv}_{O^\text{S}}, \quad \Tensor{\mathcal{M}}^u_{ijk}
=\tensor{O}^{\text{S}\,u}_{ijk},  \qquad u,v=1,2, ..., 8\,,
\nn\\
&& \qquad 
(4) \quad \text{if} \quad \mathcal{M}=\Delta,  \quad I^{uv}_\mathcal{M}
=I^{uv}_{\Delta}, \quad \Tensor{\mathcal{M}}^u_{ijk}
=\Tensor{\Delta}^u_{ijk}, \qquad u,v=1,2, ..., 10\,,
\nn
\eea
where $Q(\boldsymbol{R},\boldsymbol{x},t)$ has been defined in Eq.~(\ref{Eq:Def_Wilson_String_plus_quark}).
Integrating out the heavy-quark fields from $I^{uv}_\mathcal{M}$ we obtain 
\begin{equation}
I^{uv}_\mathcal{M} = 
\delta^3(\boldsymbol{x}_1 - \boldsymbol{y}_1)
\delta^3 (\boldsymbol{x}_2 - \boldsymbol{y}_2)
\delta^3 (\boldsymbol{x}_3 - \boldsymbol{y}_3)
\erwzero{(W^\mathcal{M}_{QQQ})^{uv}}\,,
\label{GreenQQQ}
\end{equation}
with $W^\mathcal{M}_{QQQ}$ diagrammatically represented in Fig.~\ref{fig3mat} 
and explicitly given by 
\begin{equation}
\begin{split}
(W^\mathcal{M}_{QQQ})^{uv} \equiv  \Pathorder \!\!\!\!\!\!\!\!
\sum_{ijki'j'k'rstr's't'=1}^{3}
\hspace{-5mm}
\Tensor{\mathcal{M}}^u_{ijk} \: 
\phi_{ii'}(\boldsymbol{R},\boldsymbol{x}_1,T/2)
\phi_{i'r'}(T/2,-T/2,\boldsymbol{x}_1)
\phi_{r'r}(\boldsymbol{x}_1,\boldsymbol{R},-T/2) 
\\
{} \times 
\phi_{jj'}(\boldsymbol{R},\boldsymbol{x}_2,T/2)
\phi_{j's'}(T/2,-T/2,\boldsymbol{x}_2)
\phi_{s's}(\boldsymbol{x}_2,\boldsymbol{R},-T/2) 
\\
{} \times 
\phi_{kk'}(\boldsymbol{R},\boldsymbol{x}_3,T/2)
\phi_{k't'}(T/2,-T/2,\boldsymbol{x}_3)
\phi_{t't}(\boldsymbol{x}_3,\boldsymbol{R},-T/2)  
\: \Tensor{\mathcal{M}}^{v\,*}_{rst}\,.
\end{split}
\label{WQQQ}
\end{equation}
\begin{figure}
\begin{center}
	\fmfframe(5,2)(5,5){\begin{fmfgraph*}(60,30)
	\fmfforce{(1/6w,1/3h)}{v1} 
        \fmfforce{(1/6w,1/3h)}{t1}
	\fmfforce{(1/6w,h)}{v2} 
        \fmfforce{(5/6w,h)}{v3}
	\fmfforce{(5/6w,1/3h)}{v4} 
        \fmfforce{(5/6w,1/3h)}{t2}
	\fmfforce{(0,0)}{v5} 
        \fmfforce{(4/6w,0)}{v6}
	\fmfforce{(2/6w,1/6h)}{v7} 
        \fmfforce{(w,1/6h)}{v8} 
        \fmffreeze
        \fmf{plain_arrow}{v1,v2}
	\fmf{plain_arrow}{v2,v3}
	\fmf{plain_arrow}{v3,v4} 
        \fmf{plain_arrow}{v1,v7}
	\fmf{plain_arrow}{v7,v8}
	\fmf{plain_arrow}{v8,v4} 
        \fmf{plain_arrow}{v1,v5}
	\fmf{plain_arrow}{v5,v6}
	\fmf{plain_arrow}{v6,v4} 
        \fmfdot{v1,v4}
	\fmfv{label=$\Tensor{\mathcal{M}}^{v*}_{i'j'k'}$,label.angle=60}{t1}
	\fmfv{label=$\Tensor{\mathcal{M}}^u_{ijk}$,label.angle=120}{t2}
        \fmfv{label=$Y$,label.angle=180}{v1}
	\fmfv{label=$y_1$,label.angle=180}{v2}
	\fmfv{label=$x_1$,label.angle=0}{v3}
	\fmfv{label=$X$,label.angle=30}{v4}
        \fmfv{label=$y_3$,label.angle=180}{v5}
	\fmfv{label=$x_3$,label.angle=0}{v6}
	\fmfv{label=$y_2$,label.angle=-120}{v7}
	\fmfv{label=$x_2$,label.angle=-60}{v8}
\end{fmfgraph*}}
\caption[dummy]{\label{fig3mat}
Static Wilson loop with edges $x_1 = ({\boldsymbol{x}}_1,T/2)$, $x_2 = ({\boldsymbol{x}}_2,T/2)$, 
$x_3 = ({\boldsymbol{x}}_3,T/2)$, $y_1 = ({\boldsymbol{x}}_1,-T/2)$, $y_2 = ({\boldsymbol{x}}_2,-T/2)$, 
$y_3 = ({\boldsymbol{x}}_3,-T/2)$ and insertions of the tensors  
$\Tensor{\mathcal{M}}^u_{ijk}$ and $\Tensor{\mathcal{M}}^{v\,*}_{i'j'k'}$
in $X = (\boldsymbol{R}, T/2)$ and $Y = (\boldsymbol{R},-T/2)$
respectively.}
\end{center}
\end{figure}
In the large $T$ limit, the Green functions $I_S$, $I^{uv}_{O^\text{A}}$, 
$I^{uv}_{O^\text{S}}$ and $I^{uv}_{\Delta}$ are reduced  
to the singlet, octet and decuplet propagators of pNRQCD respectively. If we
neglect subleading loop corrections to the pNRQCD side of the matching, we obtain:
\bea
&&
\hspace{-8mm} 
\lim_{T\to\infty} \erwzero{(W^\mathcal{M}_{QQQ})^{uv}} =
\lim_{T\to\infty} Z_\mathcal{M}(\bfrho,\bflambda) 
\exp\left(-iV_{\mathcal{M}}^{(0)}(\bfrho,\bflambda)\,T\right)
\nn\\
&&\hspace{-3mm} 
\times 
\erwzero{
\hspace{-5mm} 
\sum_{ijki'j'k'=1}^{3}
\hspace{-4mm} 
\Tensor{\mathcal{M}}^u_{ijk} \: 
\phi_{ii'}(T/2,-T/2,\boldsymbol{R})
\phi_{jj'}(T/2,-T/2,\boldsymbol{R})
\phi_{kk'}(T/2,-T/2,\boldsymbol{R})
\: \Tensor{\mathcal{M}}^{v\,*}_{i'j'k'}},
\eea
where $ Z_\mathcal{M}$ is a normalization factor.
At order $\als$, the result is 
\begin{align}
V_{S}^{(0)}(\bfrho,\bflambda) & = - \frac{2}{3} \,\als\,
\left(\frac{1}{|\boldsymbol{\rho}|} 
+ \frac{1}{|\boldsymbol{\lambda} + \boldsymbol{\rho}/2|} 
+ \frac{1}{|\boldsymbol{\lambda} - \boldsymbol{\rho}/2|}
\right)\,,
\label{VS3qLO}
\\
V_{O^\text{A}}^{(0)}(\bfrho,\bflambda) & =  - \frac{2}{3} \als\, 
\left( \frac{1}{|\boldsymbol{\rho}|} 
- \frac{1}{8} \frac{1}{|\boldsymbol{\lambda} + \boldsymbol{\rho}/2|}
- \frac{1}{8} \frac{1}{|\boldsymbol{\lambda} - \boldsymbol{\rho}/2|} 
\right) \,,
\label{VOALO}
\\
V_{O^\text{S}}^{(0)}(\bfrho,\bflambda) & = \frac{\als}{3} \,
\left( \frac{1}{|\boldsymbol{\rho}|} 
- \frac{5}{4} \frac{1}{|\boldsymbol{\lambda} + \boldsymbol{\rho}/2|}
- \frac{5}{4} \frac{1}{|\boldsymbol{\lambda} - \boldsymbol{\rho}/2|} 
\right) \,,
\label{VOSLO}
\\
V_{\Delta}^{(0)}(\bfrho,\bflambda) & = \frac{\als}{3} \,
\left( \frac{1}{|\boldsymbol{\rho}|} 
+ \frac{1}{|\boldsymbol{\lambda} + \boldsymbol{\rho}/2|}
+ \frac{1}{|\boldsymbol{\lambda} - \boldsymbol{\rho}/2|} 
\right) \,.
\label{VDLO}
\end{align}

\subsection{${\rm pNRQCD}$ for strongly-coupled $QQQ$ baryons}

\subsubsection{Lagrangian and Degrees of Freedom}
\begin{figure}
\makebox[-8cm]{\phantom b}
\put(0,5){\epsfxsize=8truecm \epsffile{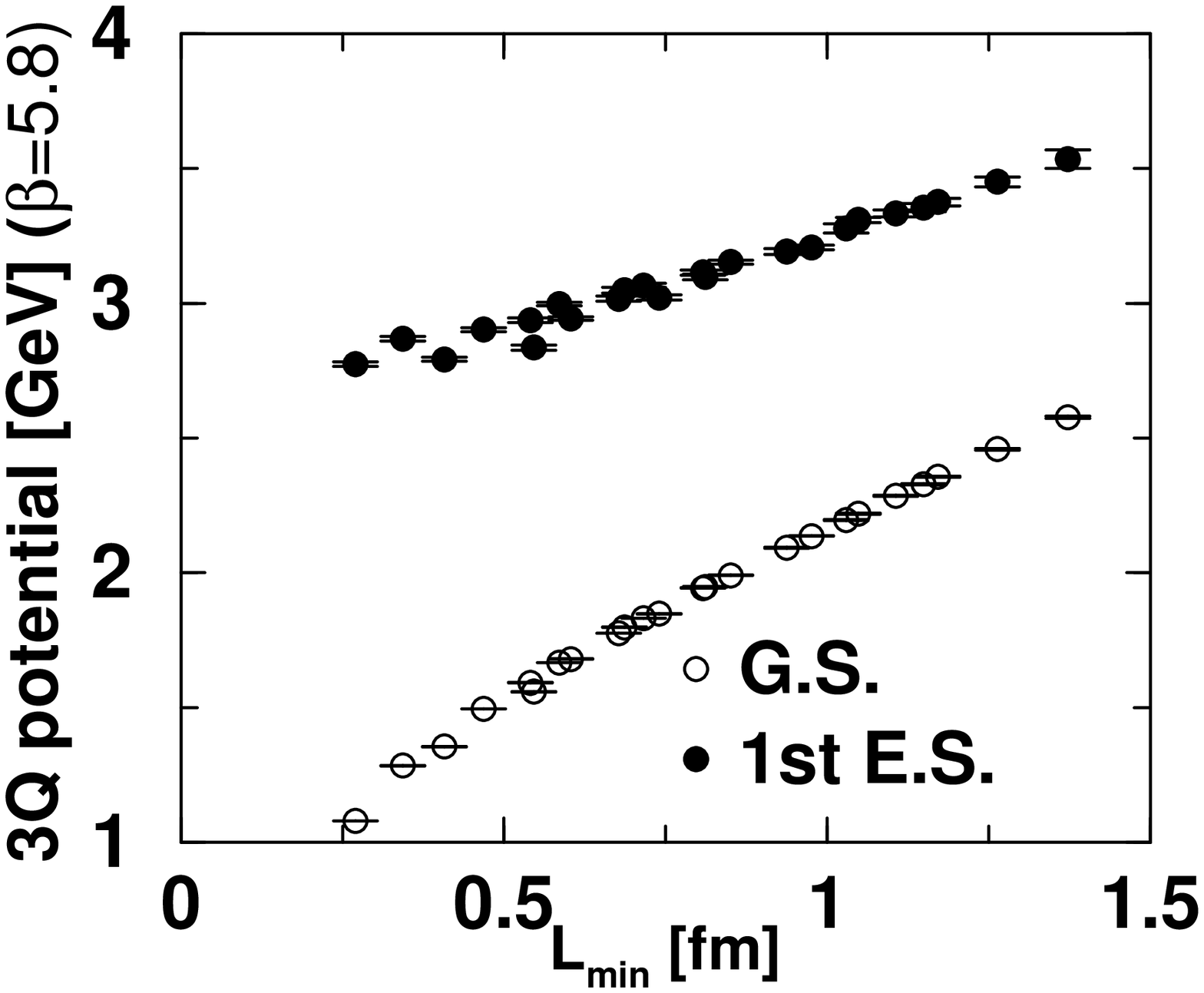}}
\put(60,34){$ E_0^{(0)}$}
\put(40,57){$ E_1^{(0)}$}
\caption{Lattice measurements of the three-quark static energies 
of the lowest state, $E_0^{(0)}$, and of the first gluonic excitation, 
$E_1^{(0)}$, as a function of  L$_\text{min}$, the minimal total length of the flux tubes 
linking the three quarks. From \cite{Takahashi:2004rw}.}
\label{figsug}
\end{figure}

In the situation in which the typical distances  $\rho$ and $\lambda$ in the baryon 
are of the order $1/\lQ$, the matching from NRQCD to pNRQCD cannot rely
on perturbation theory anymore. Also, it is more difficult 
to identify the effective degrees of freedom of pNRQCD.
Despite these difficulties, the situation appears pretty much similar to that one described 
for strongly-coupled quarkonium in \cite{Brambilla:2000gk,Brambilla:2004jw}.
From the available lattice simulations (e.g. \cite{Takahashi:2004rw}, see
Fig.~\ref{figsug}), it appears that the gluonic excitations  between three
static quarks develop an energy gap of about 1 GeV$\simg \lQ$ 
with respect to the lowest static energy. This means that all gluonic
excitations between the heavy quarks are integrated out once we go to $\hbox{pNRQCD}$.
pNRQCD in its simplest formulation, i.e. without light quark degrees 
of freedom, is, therefore, as simple as a potential model.\footnote{
The relevance of the energy gap in relation to the success of the quark model 
has also been stressed in \cite{Takahashi:2004rw}.} 
It has the three-quark singlet field $S(\bfrho,\bflambda,{\boldsymbol{R}},t)$ as the only degree of freedom 
and is described by a Lagrangian  $\mathcal{L}_\text{pNRQCD} =
\mathcal{L}_\text{pNRQCD}({\boldsymbol{R}},t)$, which reads: 
\be
\mathcal{L}_\text{pNRQCD}  = 
\int d^3\rho\,d^3\lambda \;
S^\dagger \left[ i \partial_0 
+ \frac{\boldsymbol{\nabla}^2_R}{2m_R} 
+ \frac{\boldsymbol{\nabla}^2_\rho}{2m_\rho} 
+ \frac{\boldsymbol{\nabla}^2_\lambda}{2m_\lambda} 
- V_S\right] S \,.
\ee
The potential $V_S$ may be organized in an expansion (not necessarily analytic \cite{Brambilla:2003mu})
in the inverse of the heavy-quark masses. In the following, we will consider
the matching of the $1/m$ potential, which, in the non-perturbative regime,
may, in principle, be of the same order as the static potential, and the $1/m^2$
spin-dependent potentials.

\subsubsection{Matching}
The non-perturbative matching goes as in the quarkonium case discussed 
in \cite{Brambilla:2000gk,Brambilla:2004jw} to which we refer for further
details. Here we only list some results.

The singlet static potential is given by
\be
V^{(0)}_S(\boldsymbol{\rho},\boldsymbol{\lambda})  = 
\lim_{T \rightarrow \infty} \frac{i}{T} \ln \bra{0} W^S_{QQQ} \ket{0}\,,
\ee
where $W^S_{QQQ}$ is the singlet Wilson loop defined in Eq.~(\ref{WQQQ}) and shown in Fig.~\ref{fig3mat}.
Lattice evaluations  of $V^{(0)}_S$ may be found in \cite{Alexandrou:2001ip,Takahashi:2004rw}.
A plot is shown in Fig.~\ref{figsug}.

The order $1/m$ potential is given by
\begin{equation}
V^{(1)}_S=\frac{V^{(1,1)}_S}{m} + \frac{V^{(1,3)}_S}{m_3} \,,
\label{v1}
\end{equation}
with 
\begin{align}
V^{(1,1)}_S (\boldsymbol{\rho}, \boldsymbol{\lambda}) & = 
- \frac{1}{2} \sum_{i = 1}^2 \int_{0}^{\infty} dt \,t\, 
\erw{\!\erw{g\boldsymbol{E}({\boldsymbol{x}}_i,t) \cdot g\boldsymbol{E}({\boldsymbol{x}}_i,0)}\!}^\mathcal{S}_{c,QQQ}\,,
\label{Gl:Multiplett_Potential_QQQ_1_m}\\
V^{(1,3)}_S (\boldsymbol{\rho}, \boldsymbol{\lambda}) & = 
- \frac{1}{2} \int_{0}^{\infty} dt \,t\, 
\erw{\!\erw{g\boldsymbol{E}({\boldsymbol{x}}_3,t) \cdot g\boldsymbol{E}({\boldsymbol{x}}_3,0)}\!}^\mathcal{S}_{c,QQQ}\,,
\end{align}
where the double brackets stand for the gauge field average in the presence of
a static Wilson loop of infinite time length: 
\bea
&&
\erw{\!\erw{\cdots}\!}^\mathcal{S}_{QQQ} \equiv 
\lim_{T \rightarrow \infty}
\frac{\bra{0}\cdots W^\mathcal{S}_{QQQ} \ket{0}}{\bra{0} W^\mathcal{S}_{QQQ} \ket{0}}\,,
\\
&&\erw{\!\erw{O_1(t_1) O_2(t_2)}\!}^\mathcal{S}_{c,QQQ} \equiv 
\erw{\!\erw{O_1(t_1) O_2(t_2)}\!}^\mathcal{S}_{QQQ} 
- \erw{\!\erw{O_1(t_1)}\!}^\mathcal{S}_{QQQ}
\erw{\!\erw{O_2(t_2)}\!}^\mathcal{S}_{QQQ}
\\
&&\text{with}\quad \frac{T}{2} \geq t_1 \geq t_2 \geq - \frac{T}{2}. 
\nn
\eea
As in the quarkonium case \cite{Brambilla:2000gk}, in the 
non-perturbative regime the $1/m$ potential may, in principle, be
of order $mv^2$, and therefore as important as the static potential. There are no
available lattice data for this quantity.

For the potentials responsible for the spin splittings of the heavy baryons, we obtain at order $1/m^2$:
\bea
V^{(2,\text{spin dep.})}_S &=& 
  \sum_{i=1}^3  \frac{c_S^{(i)}}{4 m_i^2} 
\bfsigma^{(i)}\cdot \left[(\bfnabla_{{\boldsymbol{x}}_i} V^{(0)}_S) \times (-i\bfnabla_{{\boldsymbol{x}}_i})\right]
\nn\\
&+&  \sum_{i,i' =1}^3 i \frac{c_F^{(i)}}{m_i m_{i'}} 
\int_0^\infty dt\,t  \, \sum_{kl=1}^3
\erw{\!\erw{g \boldsymbol{B}^k({\boldsymbol{x}}_i,t) \, g \boldsymbol{E}^l({\boldsymbol{x}}_{i'},0)}\!}^\mathcal{S}_{c,QQQ}
\bfsigma^{(i)}_k (-i\bfnabla^l_{{\boldsymbol{x}}_{i'}})
\nn\\
&-& \sum_{i> i'=1}^3 i \frac{c_F^{(i)}c_F^{(i')}}{2 m_i m_{i'}}
\int_0^\infty dt \, \sum_{kl=1}^3
\erw{\!\erw{g \boldsymbol{B}^k({\boldsymbol{x}}_i,t) \, g \boldsymbol{B}^l({\boldsymbol{x}}_{i'},0)}\!}^\mathcal{S}_{c,QQQ}
\bfsigma^{(i)}_k \bfsigma^{(i')}_l
\nn\\
&-& \sum_{i > i'=1}^3 
\left( d^{sv}_{Q_iQ_{i'}} + d^{vv}_{Q_iQ_{i'}}
\erw{\!\erw{T^{a\,(i)} T^{a\,(i')}}\!}^\mathcal{S}_{c,QQQ}
\right) \bfsigma^{(i)} \cdot \bfsigma^{(i')}
\delta^3(\boldsymbol{x}_{i} - \boldsymbol{x}_{i'})\,,
\eea 
where $T^{a\,(i)} T^{a\,(i')}$ stands for two colour matrices $T^a$ inserted at
the same time in the Wilson lines of spatial 
coordinates ${\boldsymbol{x}}_i$ and ${\boldsymbol{x}}_{i'}$ respectively, and  
the matching coefficients $d^{sv}_{Q_iQ_{i'}}$ and $d^{vv}_{Q_iQ_{i'}}$ have
been calculated in appendix \ref{appA}. 
The above expressions give at order $\als$ the well-known one-gluon exchange 
results \cite{DeRujula:1975ge}.
The spin-dependent potentials have not been calculated on the lattice yet,
differently from the quarkonium case, where such calculations have instead 
a long history \cite{Bali:2000gf}. Model dependent predictions may be found 
in \cite{Capstick:1986bm,Ford:1989ss,Brambilla:1993zw}.
It is expected that these potentials satisfy some exact relations 
due to Poincar\'e invariance of the type studied in 
\cite{Brambilla:2003nt,Brambilla:2001xk,Gromes:1984ma} for the quarkonium case.

\section{Conclusion and Outlook}
\label{secCON}
This work is a first step in the direction of a complete 
study of baryons made of two or three heavy quarks in the framework of
non-relativistic EFTs of QCD. For both types of baryons, we identify the degrees of freedom 
and write the pNRQCD Lagrangian appropriate to describe the system in the heavy-quark
sector.  In the doubly heavy baryon case this represents an update of
Ref.~\cite{Savage:di}, which, however, used a  HQET framework. 
In the case of baryons made of three heavy quarks, we also provide
non-perturbative expressions for some of the potentials.
Relevantly for both types of systems, we calculate the one-loop 
matching of the 4-quark operators of lowest dimensionality.

Several further developments are possible. For doubly heavy baryons, where 
data are already available, an important step
forward would consist in providing pNRQCD with a light-quark sector that
fully implements chiral symmetry and chiral symmetry breaking effects.
One could then study, for instance, isospin splittings and transitions 
and also address a variety of decay and production processes. 
The pursuit of such a program of phenomenological studies will, however, very much 
depend on the future of the experimental searches for these states. 

For what concerns heavy baryons made of three heavy quarks, in the absence of
a discovery, lattice studies will remain the main source of information. 
First of all, it will be important to have at least the one-loop expressions 
for the heavy-baryon static potentials $V_{S}^{(0)}$, $V_{O^\text{A}}^{(0)}$,
$V_{O^\text{S}}^{(0)}$ and $V_{\Delta}^{(0)}$ (also for $V_{T}^{(0)}$ and $V_{\Sigma}^{(0)}$). 
This may lead to a precise comparison of short-range lattice data with perturbative QCD 
in the heavy-baryon sector. At three loop, the heavy-baryon static potentials 
exhibit an ultrasoft running like in the heavy-quarkonium case \cite{Brambilla:1999qa}.
The ultrasoft running of the singlet static potential $V_{S}^{(0)}$ 
comes from the coupling with the octets $O^{\text{A}}$ and $O^{\text{S}}$. 
The leading logarithmic contribution at order $\als^4$ is
\bea
\delta V^{(0)}_S &=&  \frac{4}{9} \frac{\als}{\pi} \bflambda^2 
\left(V^{(0)}_{O^\text{A}} - V^{(0)}_{S} \right)^3 
\ln \frac{\left(V^{(0)}_{O^\text{A}} - V^{(0)}_{S} \right)^2}{4\pi\mu^2}
\nn\\
& & 
+ \frac{1}{3} \frac{\als}{\pi} \bfrho^2 
\left(V^{(0)}_{O^\text{S}} - V^{(0)}_{S} \right)^3 
\ln \frac{\left(V^{(0)}_{O^\text{S}} - V^{(0)}_{S} \right)^2}{4\pi\mu^2}
\,,
\eea
where $\mu$ is the ultrasoft factorization scale.

Let us comment on the renormalon singularities affecting the 
perturbative series of the baryonic static potentials.
These must cancel in physical observables. In the quarkonium 
case, the renormalon of the static potential cancels against 
twice the renormalon affecting the heavy-quark pole masses (see e.g. \cite{Hoang:1998nz}).
From Eq.~(\ref{VS3qLO}) one can read that 
the order $\lQ$ renormalon affecting $V_S^{(0)}$ is $3\times 1/2$ 
that one of the static potential in the quarkonium case.
Indeed, in the expression of the baryon mass it cancels against three times 
the renormalon affecting the heavy-quark pole masses.
Similarly, in the doubly heavy baryon case, from Eq.~(\ref{VTLO}) we have that 
the renormalon of order $\lQ$ affecting $V_T^{(0)}$ is $1/2$ that 
one of the static potential in the quarkonium case. 
In the expression of the baryon mass, it cancels against the renormalon 
affecting the $\bar{\Lambda}$ parameter of the HQET and the two heavy-quark masses.

Concerning the energies of gluonic excitations from three static sources, in the short-range 
they are expected to behave like the singlet potential (\ref{VS3qLO}), if they
are singlet plus glueball states, or like the octet or decuplet potentials (\ref{VOALO})-(\ref{VDLO}) 
if they are hybrid states. Only if $E_1^{(0)}$ corresponds to the first case,   
the Coulomb contribution is expected to cancel in $E_1^{(0)} - E_0^{(0)}$, 
which is the difference between the energy of the first excited state and the ground state. 
This could be in contradiction with a statement in
Ref.~\cite{Takahashi:2004rw}, where the Coulomb contribution is
said to cancel in the difference without any further specification. 
Like in the quarkonium case \cite{Brambilla:1999xf}, it is expected that the ordering of the
levels of the gluonic excitations in the short range is dictated by the correlation 
lengths of some gluonic operators. If we assume that correlation lengths of operators of higher dimensions are
suppressed and if we consider that there is a singlet channel only in $8 \otimes
8$ but not in $10 \otimes 8$, then, in the short range, the leading gluonic excitation 
is expected to come from the coupling of an octet heavy-quark state 
with a gluon field. It would be interesting to investigate if 
the first gluonic excitation shown in Fig.~\ref{figsug} is such an octet 
hybrid, and in this case what kind of octet. If it is not an octet hybrid, then 
likely it exists a lower gluonic excitation that still needs to be identified.

Finally, in the perspective of a future spectroscopy of baryons made of three
heavy quarks, it may become important to have a lattice determination of the
spin-dependent potentials. Moreover, as the history 
of quarkonium suggests, spin-dependent observables may provide an excellent insight into the
quark-confinement mechanism in the baryonic sector.

\section*{Acknowledgments}
We are grateful to Dieter Gromes for several enlightening and useful 
discussions and for  collaboration at  the initial stage of this paper. 
We thank Tom Mehen and Joan Soto for useful discussions and Randy Lewis for communications.
Two of us (N.B. and A.V.) thank the Institute for Nuclear Theory at the University of 
Washington for its hospitality and the Department of Energy for partial support during 
the last phase of completion of this work.
A.V. acknowledges the financial support obtained inside the Italian 
MIUR program  ``incentivazione alla mobilit\`a di studiosi stranieri e 
italiani residenti all'estero''.

\vfill\eject

\appendix

\section{1-loop Matching of 4-Quark Operators of Dimension 6}
\label{appA}
\begin{figure}[h]
\makebox[-11cm]{\phantom b}
\put(0,0){\epsfxsize=12truecm \epsffile{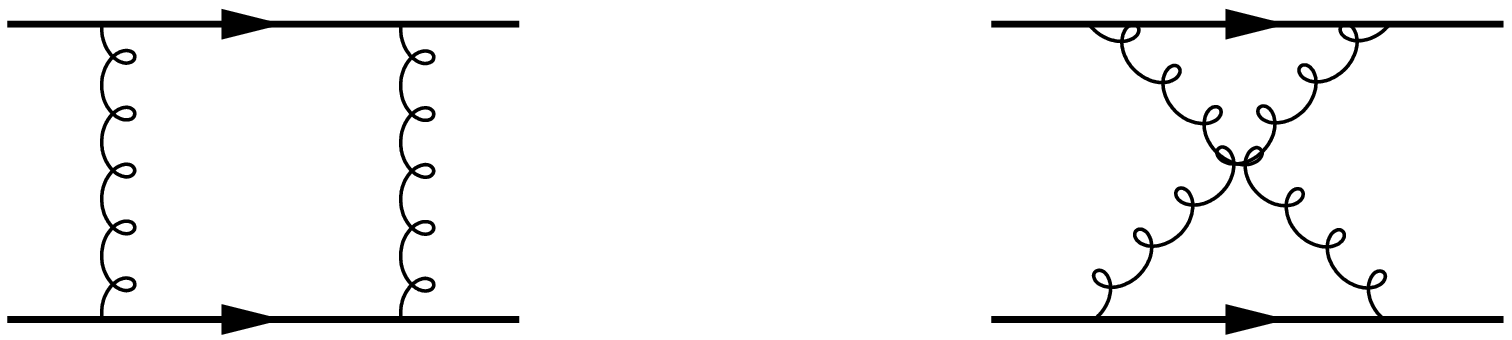}}
\caption{Feynman diagrams contributing to the 1-loop matching of the 4-quark
  operators of dimension 6.}
\label{figQQQQ}
\end{figure}

The only graphs contributing to the 1-loop matching of the 4-quark operators of dimension 6 
are displayed in Fig.~\ref{figQQQQ}. The situation is similar 
to the quark-antiquark case with different masses studied in \cite{Pineda:1998kj}.

In the case of 2 different quarks of masses $m_h$ and $m_{h'}$ ($h\neq h'$) we
obtain in the $\MS$ scheme:
\bea
d^{ss}_{Q_hQ_{h'}} &=&
   C_F \left(\frac{C_A }{ 2} -C_F \right)
    \frac{\als^2 }{ m_h^2-m^2_{h'}}
\left\{m_h^2\left(  \ln\frac{m^2_{h'} }{  \mu^2}
                   + \frac{1 }{ 3} \right)
       -
       m^2_{h'}\left(  \ln\frac{m^2_h }{  \mu^2}
                   + \frac{1 }{ 3} \right)
\right\},
\label{dss}
\\
d^{sv}_{Q_hQ_{h'}} &=& C_F \left(\frac{C_A }{ 2} -C_F \right)
   \frac{\als^2 }{ m_h^2-m^2_{h'}}
m_h m_{h'}\ln\frac{m^2_h }{ m^2_{h'}},
\label{dsv}
\\
d^{vs}_{Q_hQ_{h'}} &=&
 \frac{2 C_F \als^2 }{ m_h^2-m^2_{h'}}
  \left\{m_h^2\left( \ln\frac{m^2_{h'} }{ \mu^2}   + \frac{1 }{ 3} \right)
       - m_{h'}^2\left( \ln\frac{m^2_h }{ \mu^2}   + \frac{1 }{ 3} \right)
  \right\}
\nn\\
& & \hspace{-8mm}
- \frac{ C_A \als^2 }{ 4 (m_h^2-m^2_{h'})}
 \Biggl[
  3\left\{m_h^2\left( \ln\frac{m^2_{h'} }{ \mu^2} + \frac{1 }{ 3} \right)
       - m^2_{h'}\left(  \ln\frac{m^2_h }{ \mu^2} + \frac{1 }{ 3} \right)
  \right\}
\nn\\
&&
\quad\quad\quad\quad
  - \frac{ 1 }{ m_hm_{h'}}
   \left\{m_h^4\left( \ln\frac{m^2_{h'} }{ \mu^2} + \frac{10 }{ 3} \right)
       - m^4_{h'}\left(  \ln\frac{m^2_h }{  \mu^2} + \frac{10 }{ 3} \right) \right\}
 \Biggr],
\label{dvs}
\\
d^{vv}_{Q_hQ_{h'}}&=& \frac{2 C_F \als^2 }{ m_h^2-m_{h'}^2} m_h m_{h'}\ln\frac{m^2_h }{ m^2_{h'}}
\nn\\
& & \hspace{-8mm}
- \frac{ C_A \als^2 }{ 4 (m_h^2-m^2_{h'})}
    \Biggl[ \left\{m_h^2\left( \ln\frac{m^2_{h'} }{  \mu^2} + 5 \right)
       - m^2_{h'}\left(  \ln\frac{m^2_h }{ \mu^2} + 5 \right) \right\}
    + 3 m_h m_{h'}\ln\frac{m^2_h }{ m^2_{h'}}
    \Biggr] ,
\label{dvv}
\eea
where $C_A=N_c=3$ and $C_F=(N_c^2-1)/(2N_c)=4/3$.
For $m_h=m_{h'}=m$ the above formulas become
\bea
d^{ss}_{QQ} &=& C_F\left( \frac{C_A}{ 2}-C_F\right) \als^2 \left(  \ln\frac{m^2 }{
  \mu^2} - \frac{2 }{ 3} \right)\,,
\\
d^{sv}_{QQ}&=& C_F\left( \frac{C_A}{ 2}-C_F\right) \als^2 \,,
\\
d^{vs}_{QQ}&=& 2 C_F \als^2 \left(  \ln\frac{m^2 }{  \mu^2} - \frac{2 }{ 3} \right)
  - \frac{1 }{ 4} C_A \als^2 \left(  \ln\frac{m^2 }{  \mu^2} - \frac{23 }{ 3} \right)\,,
\\
d^{vv}_{QQ}&=& 2 C_F \als^2 - \frac{  C_A \als^2}{ 4} \left(  \ln\frac{m^2 }{\mu^2} +7 \right)
\,.
\eea
Working in $D$ dimensions, we have used the prescription 
$\epsilon_{ijk}\epsilon_{ijk} = (D-1)(D-2)(D-3)$. 
If the prescription $\epsilon_{ijk}\epsilon_{ijk} = (D-1)(D-2)$ of
\cite{Pineda:1998kj} is instead used, this amounts to changing  
$d^{vv}_{Q_hQ_{h'}} \to d^{vv}_{Q_hQ_{h'}} + C_A\als^2/2$.

\section{Group Factors}
\label{appB}

\subsection{Multiplet Tensors: $3 \otimes 3$}
\label{App:Multiplet_TensorsQQ}
The product of two triplet representations of $SU(3)$ may be decomposed 
into the sum of an antitriplet and a sextet representation: $3 \otimes 3 = \bar{3} \oplus 6$.
A possible matrix representation for the antitriplet 
($\tensor{T}^\ell_{ij}$, $\ell,i,j =  1,2,3$) and the sextet
($\Tensor{\Sigma}^\sigma_{ij}$, $\sigma =1,2, ..., 6$ and $i,j =  1,2,3$) is 
\begin{align}
&\tensor{T}^\ell_{ij}  = \frac{1}{\sqrt{2}} \epsilon_{\ell ij},
\label{Tabg}
\\
& \notag\\
&\Tensor{\Sigma}^1_{11}  = \Tensor{\Sigma}^4_{22} = \Tensor{\Sigma}^6_{33} = 1,
\notag\\
&\Tensor{\Sigma}^2_{12}  = \Tensor{\Sigma}^2_{21} = 
 \Tensor{\Sigma}^3_{13}  = \Tensor{\Sigma}^3_{31} = 
 \Tensor{\Sigma}^5_{23}  = \Tensor{\Sigma}^5_{32} = \frac{1}{\sqrt{2}}, 
\\
&\hbox{all other entries are zero.}
\notag
\end{align}
Both $\tensor{T}^\ell_{ij}$ and $\Tensor{\Sigma}^\sigma_{ij}$ are real; 
$\tensor{T}^\ell_{ij}$ is totally antisymmetric and
$\Tensor{\Sigma}^\sigma_{ij}$ totally symmetric.
They satisfy the orthogonality and normalization relations:
\be
\sum_{ij=1}^3 \tensor{T}^\ell_{ij} \, \tensor{T}^{\ell'}_{ij} =
\delta^{\ell\ell'}\,, \qquad
\sum_{ij=1}^3 \Tensor{\Sigma}^\sigma_{ij} \, \Tensor{\Sigma}^{\sigma'}_{ij} = 
\delta^{\sigma\sigma'}\,, \qquad
\sum_{ij=1}^3 \tensor{T}^\ell_{ij} \, \Tensor{\Sigma}^\sigma_{ij} = 0\,.
\ee

\subsection{Multiplet Tensors: $3 \otimes 3 \otimes 3$}
\label{App:Multiplet_TensorsQQQ}

The product of three triplet representations of $SU(3)$ may be decomposed 
into the sum of a singlet, two octet and a decuplet representation: $3 \otimes 3
\otimes 3 = 1 \oplus 8 \oplus 8 \oplus 10$.
A possible matrix representation for the singlet 
($\tensor{S}_{ijk}$, $i,j,k =  1,2,3$), the octets 
($\tensor{O}^{\text{A}a}_{ijk}$ and $\tensor{O}^{\text{S}a}_{ijk}$, $a=
1,2,...,8$, $i,j,k =  1,2,3$) and the decuplet 
($\tensor{\Delta}^{\delta}_{ijk}$, $\delta= 1,2,...,10$, $i,j,k =  1,2,3$) is 
\begin{align}
&\tensor{S}_{ijk} = \frac{1}{\sqrt{6}} \epsilon_{ijk}\,,
\\
&\notag\\
&\tensor{O}^{\text{A}\,a}_{ijk} = \frac{1}{2}\sum_{n=1}^3 \epsilon_{ijn} \lambda^a_{kn}\,,
\\
&\notag\\
&\tensor{O}^{\text{S}\,a}_{ijk} = 
\frac{1}{2\sqrt{3}} 
\sum_{n=1}^3\left( \epsilon_{jkn}\lambda^a_{in}+ \epsilon_{ikn} \lambda^a_{jn} \right)\,,
\\
&\notag\\
&\Tensor{\Delta}^1_{111} = \Tensor{\Delta}^4_{222} = \Tensor{\Delta}^{10}_{333} = 1 \,,
\notag\\
&\Tensor{\Delta}^2_{112} = \Tensor{\Delta}^2_{121} = \Tensor{\Delta}^2_{211} = 
\Tensor{\Delta}^3_{122}  = \Tensor{\Delta}^3_{212} = \Tensor{\Delta}^3_{221} =
\frac{1}{\sqrt{3}} \,,
\notag\\
&\Tensor{\Delta}^5_{113} = \Tensor{\Delta}^5_{131} = \Tensor{\Delta}^5_{311} = 
\Tensor{\Delta}^7_{223}  = \Tensor{\Delta}^7_{232} = \Tensor{\Delta}^7_{322} = 
\frac{1}{\sqrt{3}} \,,
\\
&\Tensor{\Delta}^8_{133} = \Tensor{\Delta}^8_{313} = \Tensor{\Delta}^8_{331} = 
\Tensor{\Delta}^9_{233}  = \Tensor{\Delta}^9_{323} = \Tensor{\Delta}^9_{332} = 
\frac{1}{\sqrt{3}} \,,
\notag\\
&\Tensor{\Delta}^6_{123} = \Tensor{\Delta}^6_{132}  = \Tensor{\Delta}^6_{213}= 
\Tensor{\Delta}^6_{231}  = \Tensor{\Delta}^6_{312}  = \Tensor{\Delta}^6_{321}  = 
\frac{1}{\sqrt{6}} \,,
\notag\\
&\hbox{all other entries are zero,}
\notag
\end{align}
where $\lambda^a$ are the Gell-Mann matrices. $\tensor{S}_{ijk}$ and
$\tensor{\Delta}^{\delta}_{ijk}$ are real; $\tensor{S}_{ijk}$ is totally
antisymmetric and $\tensor{\Delta}^{\delta}_{ijk}$ totally symmetric.
The octets $\tensor{O}^{\text{A}a}_{ijk}$ and $\tensor{O}^{\text{S}a}_{ijk}$ 
have been chosen to be respectively antisymmetric and symmetric in the first 
two indices. The matrices satisfy the orthogonality and normalization relations:
\bea
&&\hspace{-4mm}
\sum_{ijk=1}^3 \tensor{S}_{ijk} \, \tensor{S}_{ijk}  = 1\,, 
\;
\sum_{ijk=1}^3 \tensor{O}^{\text{A}\,a\,*}_{ijk} \, \tensor{O}^{\text{A}\,a'}_{ijk}  = \delta^{aa'}\,, 
\; 
\sum_{ijk=1}^3 \tensor{O}^{\text{S}\,a\,*}_{ijk} \,\tensor{O}^{\text{S}\,a'}_{ijk}  = \delta^{aa'}\,,
\; 
\sum_{ijk=1}^3 \Tensor{\Delta}^{\delta}_{ijk} \, \Tensor{\Delta}^{\delta'}_{ijk}  = \delta^{\delta\delta'}\,,
\nn\\
&& \hspace{-4mm}
\sum_{ijk=1}^3 \tensor{S}_{ijk} \, \tensor{O}^{\text{A}\,a}_{ijk}  
= \sum_{ijk=1}^3 \tensor{S}_{ijk} \, \tensor{O}^{\text{S}\,a}_{ijk}  
= \sum_{ijk=1}^3 \tensor{S}_{ijk} \,  \Tensor{\Delta}^{\delta}_{ijk} 
= 0\,,
\\
&& \hspace{-4mm}
\sum_{ijk=1}^3 \tensor{O}^{\text{A}\,a\,*}_{ijk} \,\tensor{O}^{\text{S}\,a'}_{ijk}  
=\sum_{ijk=1}^3 \tensor{O}^{\text{A}\,a\,*}_{ijk} \, \Tensor{\Delta}^{\delta}_{ijk} 
= \sum_{ijk=1}^3 \tensor{O}^{\text{S}\,a\,*}_{ijk}\,
\Tensor{\Delta}^{\delta}_{ijk} 
=0\,.
\nn
\eea

\subsection{$SU(3)$ Representations}
\label{App:Group_rep}
Here we list our choice of matrix representations for the $SU(3)$ 
generators in the $3$, $\bar{3}$, $6$, $8$, $10$ representations: 
\begin{align}
T^a \equiv T^a_3 & \equiv \frac{\lambda^a}{2}\,, 
\label{Gl:Def_SU3_Darstellung_3}\\
T^a_{\bar{3}} & \equiv - \frac{\lambda^{a\transpose}}{2}\,,
\label{Gl:Def_SU3_Darstellung_3bar}\\
(T^a_6)_{\sigma\sigma'} & \equiv \sum_{ijk=1}^3 
\Tensor{\Sigma}^\sigma_{ij} \lambda^a_{jk}\Tensor{\Sigma}^{\sigma'}_{ki},
\qquad\qquad\;\;
\sigma,\sigma'=1,2,\dots,6\,,
\label{Gl:Def_SU3_Darstellung_6}\\
(T^a_8)_{bc} & \equiv if^{bac},\qquad\qquad\qquad\qquad\quad
 b,c=1,2,\dots,8\,,
\label{Gl:Def_SU3_Darstellung_8}\\
(T^a_{10})_{\delta\delta'} & \equiv \frac{3}{2} \sum_{ii'jk=1}^3 
\Tensor{\Delta}^\delta_{ijk}\lambda^a_{ii'} \Tensor{\Delta}^{\delta'}_{i'jk},
\qquad \delta,\delta'=1,2,\dots,10\,,
\label{Gl:Def_SU3_Darstellung_10}
\end{align}
where $a=1,2,\dots,8$.

\section{Spin States}
\label{appC}
Let us consider a meson made by a heavy antiquark $Q$ and a light quark $q$.
We denote by $|P^*;{\bf S}^z_{\bar{Q}q}\rangle$ the lowest $S_{\bar{Q}q}=1$
states (${\bf S}^z_{\bar{Q}q}=1,0,-1$), 
and by $|P;0\rangle$ the lowest $S_{\bar{Q}q}=0$ state. 
The heavy antiquark content of the states may be made explicit by writing:
\bea
\vert P^*; 1 \rangle &=& 
\int d^3R \; Q^\dagger_+({\boldsymbol{R}}) \vert {\bf S}_l^z =1/2 \rangle,
\label{spinstat0}
\\
\vert P^*; 0 \rangle &=& 
\int d^3R \; \frac{1}{ \sqrt{2}} \left(
 Q^\dagger_+({\boldsymbol{R}}) \vert {\bf S}_l^z =-1/2 \rangle
+ Q^\dagger_-({\boldsymbol{R}}) \vert {\bf S}_l^z =1/2 \rangle  \right), 
\\
\nn
\eea
\bea
\vert P^*; -1 \rangle &=& 
\int d^3R \; Q^\dagger_-({\boldsymbol{R}}) \vert {\bf S}_l^z =-1/2 \rangle,  
\\
\vert P; 0 \rangle &=& 
\int d^3R \; \frac{1}{ \sqrt{2}} \left(
Q^\dagger_+({\boldsymbol{R}}) \vert {\bf S}_l^z =-1/2 \rangle 
-Q^\dagger_-({\boldsymbol{R}}) \vert {\bf S}_l^z =1/2 \rangle  \right).
\label{spinstatn}
\eea

In the case of the lowest doubly heavy baryon states, 
we denote by $|\Xi^*;{\bf S}^z_{QQq}\rangle$ the $S_{QQq}=3/2$ states 
(${\bf S}^z_{QQq}=\pm 3/2,\pm 1/2$), and by $|\Xi;\pm 1/2\rangle$ the $S_{QQq}=1/2$ states. 
The heavy antitriplet content of the states may be made explicit by writing:
\bea
\vert \Xi^*; 3/2 \rangle &=& 
\int d^3R d^3r \; \varphi_{QQ}({\boldsymbol{r}}) \, T^\dagger_+({\boldsymbol{r}},{\boldsymbol{R}}) 
\vert {\bf S}_l^z =1/2 \rangle,
\label{spinstatnqq0}
\\
\vert \Xi^*; 1/2 \rangle &=& 
\int d^3R d^3r \; \varphi_{QQ}({\boldsymbol{r}}) \, \left( 
\sqrt{\frac{1}{3}} T^\dagger_+({\boldsymbol{r}},{\boldsymbol{R}}) \vert {\bf S}_l^z = - 1/2 \rangle
\right.
\nn\\
&& \qquad
\left. 
+ \sqrt{\frac{2}{3}} T^\dagger_0({\boldsymbol{r}},{\boldsymbol{R}}) \vert {\bf S}_l^z = 1/2 \rangle
\right),
\\
\vert \Xi^*; -1/2 \rangle &=& 
\int d^3R d^3r \; \varphi_{QQ}({\boldsymbol{r}}) \, \left( 
\sqrt{\frac{2}{3}} T^\dagger_0({\boldsymbol{r}},{\boldsymbol{R}}) \vert {\bf S}_l^z =  -1/2 \rangle
\right.
\nn\\
&& \qquad
\left. 
+ \sqrt{\frac{1}{3}} T^\dagger_-({\boldsymbol{r}},{\boldsymbol{R}}) \vert {\bf S}_l^z =1/2 \rangle
\right),
\\
\vert \Xi^*; -3/2 \rangle &=& 
\int d^3R d^3r \; \varphi_{QQ}({\boldsymbol{r}})\, T^\dagger_-({\boldsymbol{r}},{\boldsymbol{R}}) \vert {\bf S}_l^z = - 1/2 \rangle,
\\
\vert \Xi; 1/2 \rangle &=& 
\int d^3R d^3r \; \varphi_{QQ}({\boldsymbol{r}}) \, \left( 
\sqrt{\frac{2}{3}} T^\dagger_+({\boldsymbol{r}},{\boldsymbol{R}}) \vert {\bf S}_l^z = - 1/2 \rangle
\right.
\nn\\
&& \qquad
\left. 
- \sqrt{\frac{1}{3}} T^\dagger_0({\boldsymbol{r}},{\boldsymbol{R}}) \vert {\bf S}_l^z = 1/2 \rangle
\right),
\\
\vert \Xi; -1/2 \rangle &=& 
\int d^3R d^3r \; \varphi_{QQ}({\boldsymbol{r}}) \, \left( 
\sqrt{\frac{1}{3}} T^\dagger_0({\boldsymbol{r}},{\boldsymbol{R}}) \vert {\bf S}_l^z = -1/2 \rangle
\right.
\nn\\
&& \qquad
\left. 
- \sqrt{\frac{2}{3}} T^\dagger_-({\boldsymbol{r}},{\boldsymbol{R}}) \vert {\bf S}_l^z =1/2 \rangle
\right),
\label{spinstatnqq}
\eea
where 
\be
\int d^3r \; \varphi_{QQ}^*({\boldsymbol{r}}) \varphi_{QQ}({\boldsymbol{r}}) =1.
\ee

\end{fmffile}
\end{document}